\documentclass[reprint, amsmath,amssymb, aps, superscriptaddress]{revtex4-2}
\usepackage{b26_style}
\usepackage[yyyymmdd,hhmmss]{datetime}

\newcommand{\CnNOTe}{\textrm{C}_\textrm{n} \textrm{NOT}_\textrm{e}}
\newcommand{\fifteenN}{^{15}\textrm{N}}
\newcommand{\thirteenC}{^{13}\textrm{C}}

\DeclareUnicodeCharacter{2009}{\,}
\DeclareSIUnit\gauss{G}

\begin{document}

\title{
Programmable Quantum Processors based on Spin Qubits with Mechanically-Mediated Interactions and Transport}

\author{F. Fung}
\affiliation{Department of Physics, Harvard University, Cambridge, Massachusetts, 02138, USA}

\author{E. Rosenfeld}
\thanks{Currently at AWS Center for Quantum Computing, Pasadena, California, 91106, USA}
\affiliation{Department of Physics, Harvard University, Cambridge, Massachusetts, 02138, USA}

\author{J. D. Schaefer}
\affiliation{Department of Physics, Harvard University, Cambridge, Massachusetts, 02138, USA}

\author{A. Kabcenell}
\affiliation{Department of Physics, Harvard University, Cambridge, Massachusetts, 02138, USA}

\author{J. Gieseler}
\affiliation{Department of Physics, Harvard University, Cambridge, Massachusetts, 02138, USA}
\thanks{Current address: IAV GmbH DigitalLab.}

\author{\\ T. X. Zhou}
\thanks{Currently at Northrop Grumman Mission Systems, Linthicum, Maryland, 21090, USA}
\affiliation{Department of Physics, Harvard University, Cambridge, Massachusetts, 02138, USA}
\affiliation{Harvard John A. Paulson School of Engineering and Applied Sciences, Harvard University, Cambridge, Massachusetts, 02138, USA}
\affiliation{Massachusetts Institute of Technology, Cambridge, Massachusetts, 02139, USA}

\author{T. Madhavan}
\affiliation{Harvard John A. Paulson School of Engineering and Applied Sciences, Harvard University, Cambridge, Massachusetts, 02138, USA}

\author{N. Aslam}
\affiliation{Department of Physics, Harvard University, Cambridge, Massachusetts, 02138, USA}
\affiliation{Institute of Condensed Matter Physics, Technische Universit\"{a}t Braunschweig, Braunschweig, Germany}

\author{A. Yacoby}
\affiliation{Department of Physics, Harvard University, Cambridge, Massachusetts, 02138, USA}

\author{M. D. Lukin}
\email{lukin@physics.harvard.edu}
\affiliation{Department of Physics, Harvard University, Cambridge, Massachusetts, 02138, USA}

%\date{\today~at~\currenttime}

\begin{abstract} 
Solid state spin qubits are promising candidates for quantum information processing, but controlled interactions and entanglement in large, multi-qubit systems are currently difficult to achieve. We describe a method for programmable control of multi-qubit spin systems, in which individual nitrogen-vacancy (NV) centers in diamond nanopillars are coupled to magnetically functionalized silicon nitride mechanical resonators in a scanning probe configuration. Qubits can be entangled via interactions with nanomechanical resonators while programmable connectivity is realized via mechanical transport of qubits in nanopillars. To demonstrate the feasibility of this approach, we characterize both the mechanical properties and the magnetic field gradients around the micromagnet placed on the nanobeam resonator. Furthermore, we show coherent manipulation and mechanical transport of a proximal spin qubit by utilizing nuclear spin memory, and use the NV center to detect the time-varying magnetic field from the oscillating micromagnet, extracting a spin-mechanical coupling of \SI{7.7(9)}{\Hz}. With realistic improvements the high-cooperativity regime can be reached, offering a new avenue towards scalable quantum information processing with spin qubits.

\end{abstract}

\maketitle

%%%%%%%%%%%%%%%%%%%%%%%%%%%%%%%%%%%%%%%%%%%%%%%%%%
\paragraph*{Introduction.}
%%%%%%%%%%%%%%%%%%%%%%%%%%%%%%%%%%%%%%%%%%%%%%%%%%

Isolated spin defects in the solid state, such as nitrogen vacancy (NV) centers in diamond, have long been considered as promising candidates for quantum information processing, owing to their extended coherence times even at elevated temperatures \cite{balasubramanian2009ultralong, maurer2012room,widmann2015coherent,anderson2022five,stas2022robust}. While small spin registers have been realized using coupled electronic and 
%nitrogen vacancy (NV) centers 
nuclear spins \cite{dutt2007quantum, abobeih2018one, bradley2019ten}, such demonstrations rely on magnetic dipole-dipole interactions, which limit the distance between spins to tens of nanometers. The short-range nature of these interactions and imprecision of defect fabrication at these length scales make it challenging to control systems containing large arrays of spin qubits. 

Several approaches are currently being explored to address this challenge, including long-range entanglement based on photonic \cite{bhaskar2020experimental,hermans2022qubit} and mechanical systems \cite{rabl_quantum_2010,kuzyk2018scaling}. In particular, nanomechanical resonators have been proposed as a mesoscopic interface between distant and otherwise isolated spin qubits. Such a hybrid quantum system can be realized by combining electronic spins with magnetically functionalized mechanical resonators \cite{arcizet_single_2011, bennett_measuring_2012, kolkowitz_coherent_2012, nichol_nanomechanical_2012, rugar_single_2004, teissier_strain_2014, pigeau_observation_2015, lee_strain_2016, meesala_enhanced_2016, wrachtrup_cantilever_2020, gieseler_single-spin_2020, maity_coherent_2020}. Mechanical resonators can be engineered to have very high quality factors with flexible, compact geometric realizations, and feature low crosstalk relative to their electromagnetic counterparts \cite{chu2020perspective}. Using mechanical modes as a quantum transducer, distant spin qubits can be entangled deterministically, even when the mechanical mode is in a thermal, highly excited state \cite{rabl_quantum_2010, schuetz_high-fidelity_2017, rosenfeld_efficient_2020}. Furthermore,  spin qubits can be used to cool the mechanical resonator to its ground state \cite{rabl2009strong,rabl_cooling_2010} and subsequently prepare non-Gaussian states of motion \cite{oconnell_quantum_2010}. 
Despite these intriguing proposals, realizing the necessary strong coupling between mechanical systems and individual spin qubits is a challenging task, requiring deterministic positioning of spin qubits in close proximity to magnetized mechanical resonators. Moreover, even though transducers extend the spin-spin interaction range, the system connectivity remains local, limiting its programmability and scalability. 

\begin{figure}
\includegraphics[width=\columnwidth]{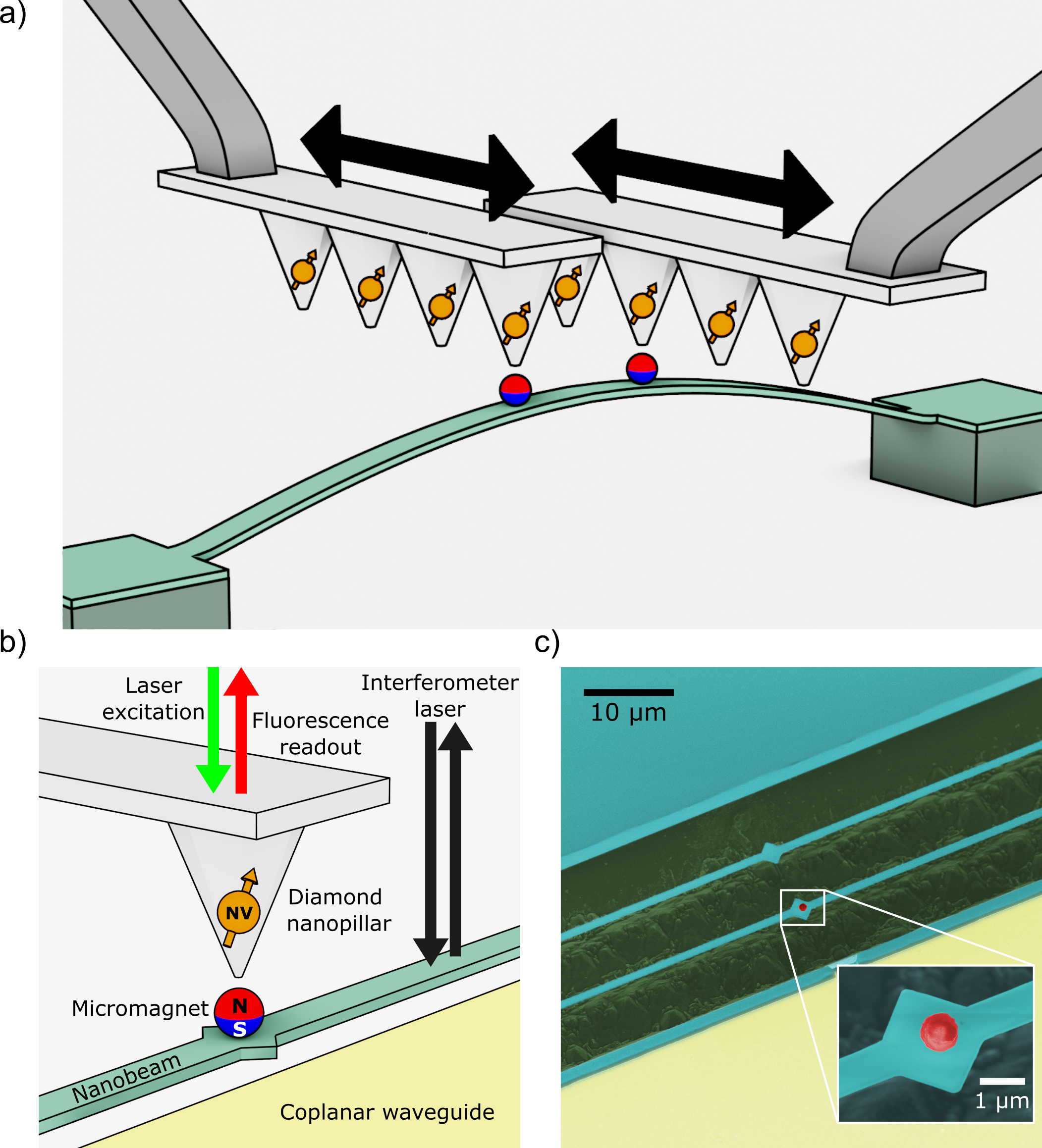}
\caption{\textrm{(a)} Conceptual diagram for programmable interactions within arrays of nanopillars containing spin qubits. Spin qubits (orange) can be micro-manipulated (black arrows) to interact with a common mechanical mode of a magnetically-functionalized resonator (turquoise), enabling mechanically-mediated spin-spin interactions. (b) Our current experimental realization: a diamond nanopillar containing a single NV center at its tip is positioned above an NdFeB spherical micromagnet at the center of a doubly clamped silicon nitride nanobeam. The nanobeam is fabricated on the same chip as a coplanar waveguide which facilitates coherent microwave control of the NV center's electronic spin. The chip is mounted on a 3-axis nanopositioner stack, while the diamond nanopillar is kept fixed. \textrm{(c)} False-color scanning electron microscope image of the nanobeam (turquoise lines, two shown) with a micromagnet (red sphere) placed on a pad at the antinode of motion. The gold area at the bottom of the image is the coplanar waveguide for microwave delivery. Inset: zoom-in image of the micromagnet.}
\label{fig:setup}
\end{figure}

In this Letter, we introduce a novel platform for realizing programmable interactions between distant spin qubits. The key idea of our architecture is illustrated in Fig.~\ref{fig:setup}(a). In our approach, individual NV centers in diamond nanopillars are coupled to silicon nitride nanobeam mechanical resonators in a scanning probe geometry. A micromagnet attached to the nanobeam provides the magnetic field gradient for the spin-mechanical coupling. Ultra-high quality factors $Q>10^9$ have been demonstrated in silicon nitride mechanical resonators through a combination of techniques such as soft-clamping, dissipation dilution, and strain engineering \cite{tsaturyan_ultracoherent_2017, ghadimi_elastic_2018, bereyhi_clamp-tapering_2019, grob_fractal_2022,kippenberg_fractal_2021}. At the same time, the nanoscale footprint of the diamond nanopillar enables small separations between the NV center and the micromagnet \cite{zhou_scanning_2017, maletinsky_robust_2012}, providing access to high magnetic field gradients required for large spin-mechanical coupling. Remarkably, spin qubits confined in nanopillars can be moved mechanically in and out of the near-field of the magnetized resonators. Moreover, they can be transported across relatively long (10-100 \SI{}{\micro\meter}) distances, enabling non-local connectivity between distant qubits \cite{dolev_transport, dhordjevic2021entanglement,mandel2003coherent,pino2021demonstration,monroe2014large,cirac2000scalable}. Further improvements in coherence time can be obtained by making use of a nuclear spin quantum memory. Since the latter is less sensitive to magnetic fields, such storage can be used for long-distant qubit transport even in the presence of proximal magnetic gradients, enabling reconfigurable quantum processing architecture similar to that demonstrated recently for neutral atom array qubits \cite{dolev_transport}.

To demonstrate the feasibility of this approach, we first perform scanning magnetometry with the nanopillar to characterize the magnetic field and field gradients around a micromagnet. Subsequently, as a proof-of-principle demonstration of coherent transport, we store coherent information inside the NV center's intrinsic $\fifteenN$ nuclear spin, and show that the spin coherence is not affected by movement over $\SI{1.7(2)}{\micro\meter}$ near the micromagnet. Finally, by measuring the mechanical motion interferometrically and using the NV center to detect the time-varying magnetic field from the oscillating micromagnet, we extract the single-phonon spin-mechanical coupling strength of $\SI{7.7(9)}{\hertz}$. With realistic improvements to our quality factors and further reduction in the magnet-NV center distance, the coherent coupling regime is within reach.

%%%%%%%%%%%%%%%%%%%%%%%%%%%%%%%%%%%%%%%%%%%%%%%%%%
\paragraph*{Experimental setup and static magnetic field characterization.}
%%%%%%%%%%%%%%%%%%%%%%%%%%%%%%%%%%%%%%%%%%%%%%%%%%

Our experimental platform consists of a scanning probe setup, where a diamond nanopillar containing a single NV center is positioned near a micromagnet placed at the center of the nanobeam (Fig.~\ref{fig:setup}(b, c)). In addition to improving the optical collection efficiency, the nanopillar has a small surface area at its apex, allowing for nanoscale magnet-NV center distances \cite{zhou_scanning_2017, maletinsky_robust_2012,angle_etch}.

\begin{figure}
\includegraphics[width = .98\columnwidth]{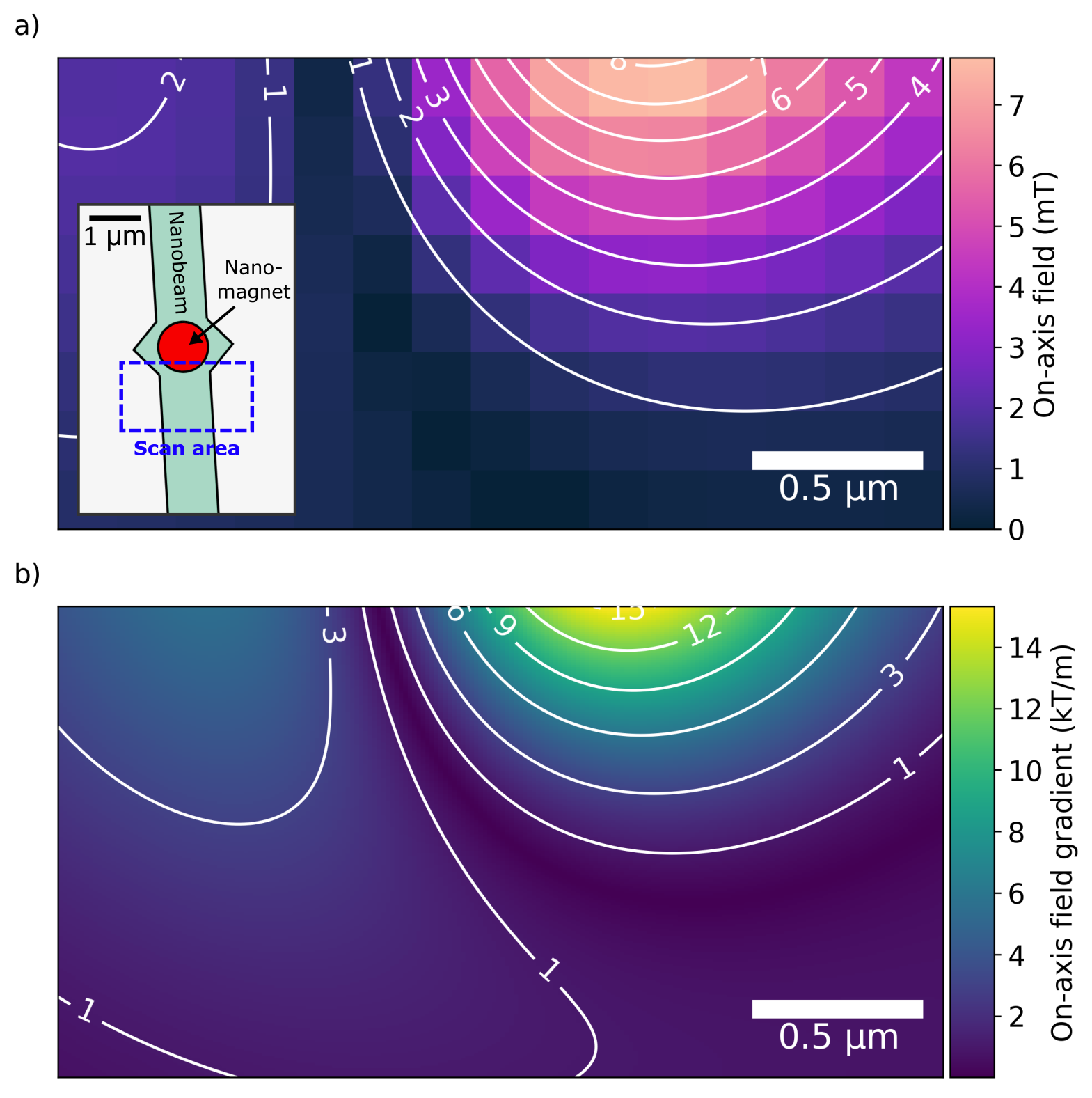}
\caption{Magnetic field imaging of a micromagnet. \textrm{(a)} Magnetic field in an area around the magnet, with a vertical NV-magnet separation of about \SI{1}{\micro\meter}. For each position of the diamond nanopillar, we measure the electron spin resonance (ESR) frequencies, from which we extract the magnetic field along the NV center's axis \cite{supp}. The field from the spherical micromagnet is well-approximated by a dipole model, with the fit (contours) deviating from the measured fields by no more than \SI{3}{\gauss} at any point. The orientations of the NV center and magnet are consistent with the fabrication process of the nanopillar and magnetization direction respectively. These measurements were performed in air at room temperature, without any external driving of the mechanical resonator. Inset: Diagram showing the scanning area around the micromagnet on the nanobeam. \textrm{(b)} Reconstructed magnetic field gradients based on fit to dipole model. The maximum gradient is $\sim\SI{1.5e4}{\tesla/\meter}$.}
\label{fig:field_scan} 
\end{figure}

The presence of a magnetic field perpendicular to the NV center quantization axis limits NV spin readout contrast, photoluminesence intensity, and coherence time $T_{2,e}$ in a natural-abundance $\thirteenC$ diamond \cite{stanwix_coherence_2010}. In order to align the magnetic field and characterize the field distribution, we scan the micromagnet with respect to the diamond nanopillar with a 3-axis stack of piezoelectric nanopositioners. At each position, we optically measure the electron spin resonance (ESR) and calculate the magnetic field along the NV center quantization axis \cite{supp}; an example map of the magnetic field around a micromagnet is shown in Fig.~\ref{fig:field_scan}(a). With our current smallest micromagnet-NV distance of $\sim\SI{1.0}{\micro\meter}$, we estimate gradients exceeding \SI{1.5e4}{\tesla/\meter}, corresponding to an expected single-phonon spin-mechanical coupling strength of $\lambda/2\pi \sim 5\,\SI{}{\hertz}$ (Fig. \ref{fig:field_scan}(b)). 

\paragraph*{Preservation of spin coherence while moving in a magnetic field gradient.}

\begin{figure}
\centering
\includegraphics[width=.98\columnwidth]{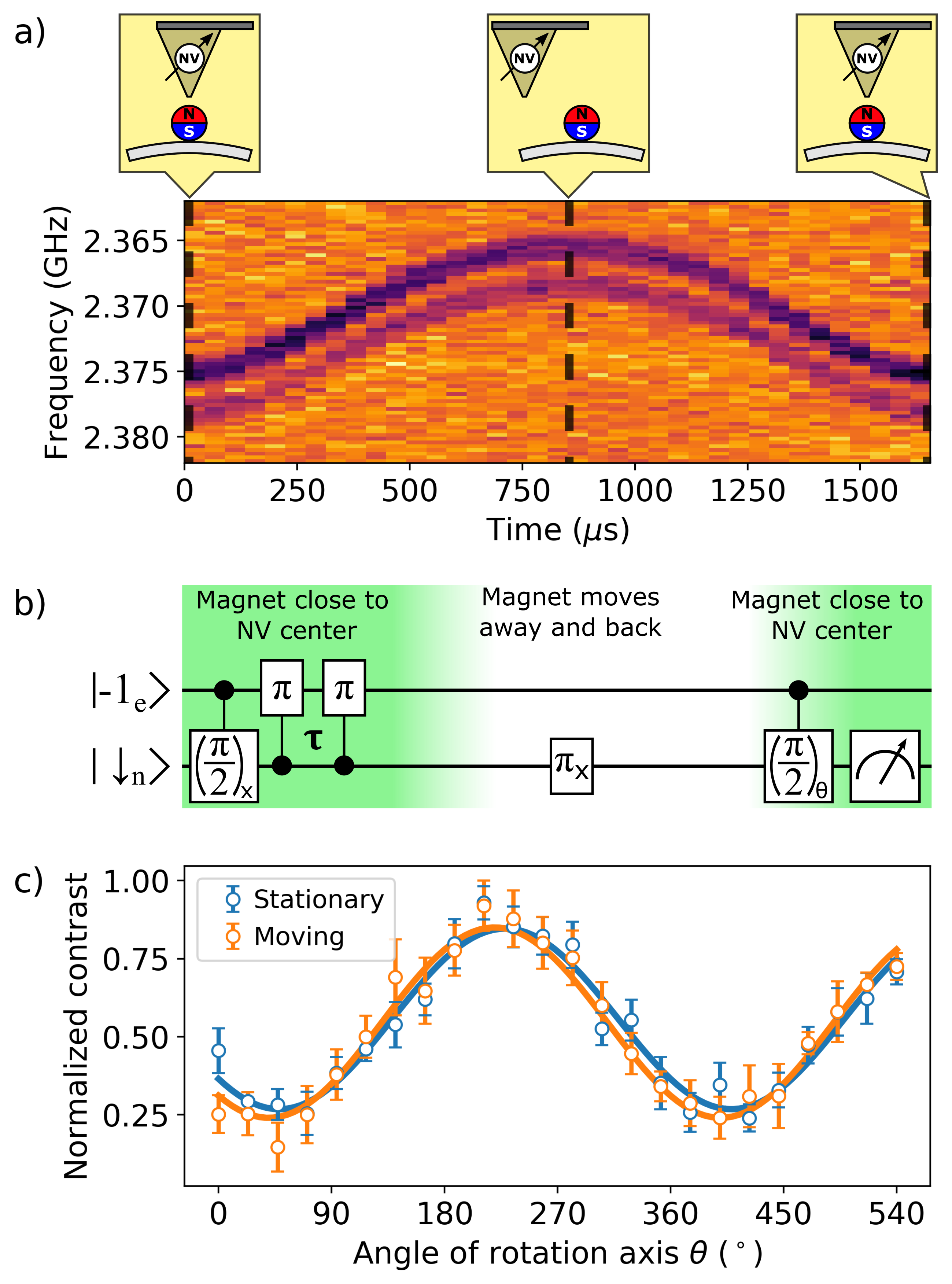}
\caption{Preservation of spin coherence while moving in a magnetic field gradient. {\rm(a)} The nanobeam along with the micromagnet is moved \SI{1.7(2)}{\micro\meter} from the NV center and then back to its original position over \SI{1.7}{\milli\second}. Pulsed ESR measurements at different times during the movement sequence show the changing field from the moving micromagnet. The \SI{3}{\mega\hertz} hyperfine splitting from the NV center's intrinsic $\fifteenN$ nuclear spin is clearly visible; an additional hyperfine splitting from a nearby $\thirteenC$ nuclear spin is not shown here due to the microwave pulse duration used \cite{supp}. {\rm(b)} Pulse sequence used to demonstrate storage and retrieval of coherent information, synchronized with the movement sequence shown in {\rm(a)}. {\rm(c)} Fixing the phase accumulation time $\tau = \SI{900}{\nano\second}$, we measure the coherence of the nuclear spin at the end of the movement sequence by varying the rotation axis angle $\theta$ of the final $\pi/2$-pulse, for both cases where the micromagnet is moved (orange) and kept stationary (blue).}
\label{fig:coherence} 
\end{figure}

Next, to investigate whether the spin coherence can be maintained during qubit transport, we perform a proof-of-principle experiment in which we move the micromagnet \SI{1.7(2)}{\micro\meter} away from the diamond nanopillar and then back to its original position. Pulsed ESR measurements at different times during the movement sequence (Fig.~\ref{fig:coherence} (a)) reveal a significant change in the magnetic field environment, as evidenced by a shift of $\SI{9.8(1)}{\mega\hertz}$ in the ESR frequency \cite{supp}.

We demonstrate the preservation of spin coherence while moving in a magnetic field gradient, using the pulse sequence in Fig.~\ref{fig:coherence}(b) synchronized with the movement sequence in Fig.~\ref{fig:coherence}(a). Since the total movement time of \SI{1.7}{\milli\second} is significantly longer than the electronic spin coherence time of \SI{0.95(4)}{\milli\second} \cite{supp}, we use 
the NV center's intrinsic $\fifteenN$ nuclear spin as a quantum memory 
\cite{pfender2017nonvolatile, zaiser2016enhancing} to enable the mechanical qubit transport. 

Specifically, in our demonstration the electron and $\fifteenN$ nuclear spin are first initialized in a two-qubit register $\ket{-1}_e \otimes \ket{\downarrow}_n$ \cite{supp}, followed by a $\pi/2$-pulse which puts the $\fifteenN$ nuclear spin in a superposition $\ket{-1}_e \otimes \left( \ket{\downarrow} + \ket{\uparrow} \right )_n$. 
Subsequently, we apply a $\CnNOTe$ gate which fully entangles the electron-nucleus pair
$-\ket{0}_e \ket{\downarrow}_n + \ket{-1}_e \ket{\uparrow}_n$.

During the subsequent free evolution time $\tau$, the entangled electron-nucleus pair accumulates a phase $\phi(\tau)$. For the particular NV center in our measurements, hyperfine interactions with a nearby $\thirteenC$ nuclear spin lead to phase accumulation at a rate of $\sim \SI{0.9}{\mega\hertz}$. A second $\CnNOTe$ gate then disentangles the electron-nuclear pair and the phase information $\phi(\tau)$ is now entirely stored in the $^{15}\textrm{N}$ nuclear spin
$\ket{-1}_e \otimes \left( -\ket{\downarrow} + e^{i \phi(\tau)} \ket{\uparrow} \right )_n$.

As shown in Fig.~\ref{fig:coherence}(a), the field at the nanopillar changes significantly during the movement sequence, leading to an additional phase accumulation on the $\fifteenN$. We eliminate this additional phase by applying a $\pi$-pulse on the $^{15}\textrm{N}$ near the middle of the movement sequence \cite{supp}. Finally, a $\pi/2$-pulse at the end of the movement sequence converts the stored phase information $\phi(\tau)$ into the probability of finding the $\fifteenN$ in either $\ket{\downarrow}_n$ or $\ket{\uparrow}_n$, which can be measured with a boost in signal-to-noise ratio (SNR) using repetitive readout \cite{jiang2009repetitive}.

By fixing $\tau = 900\,\SI{}{\nano\second} < T_{2,e}^*$ and varying the rotation axis angle of the final $\pi/2$-pulse, we can quantify the spin coherence preservation. The results, shown in \ref{fig:coherence}(c) demonstrate that the normalized contrasts for cases where the micromagnet is moved (orange) and kept stationary (blue) are $0.61(3)$ and $0.57(3)$ respectively, indicating that the nuclear spin coherence is unaffected by the significant change in magnetic field.

\paragraph*{Mechanical motion and single-phonon coupling strength.}

\begin{figure}
\centering
\includegraphics[width=.97\columnwidth]{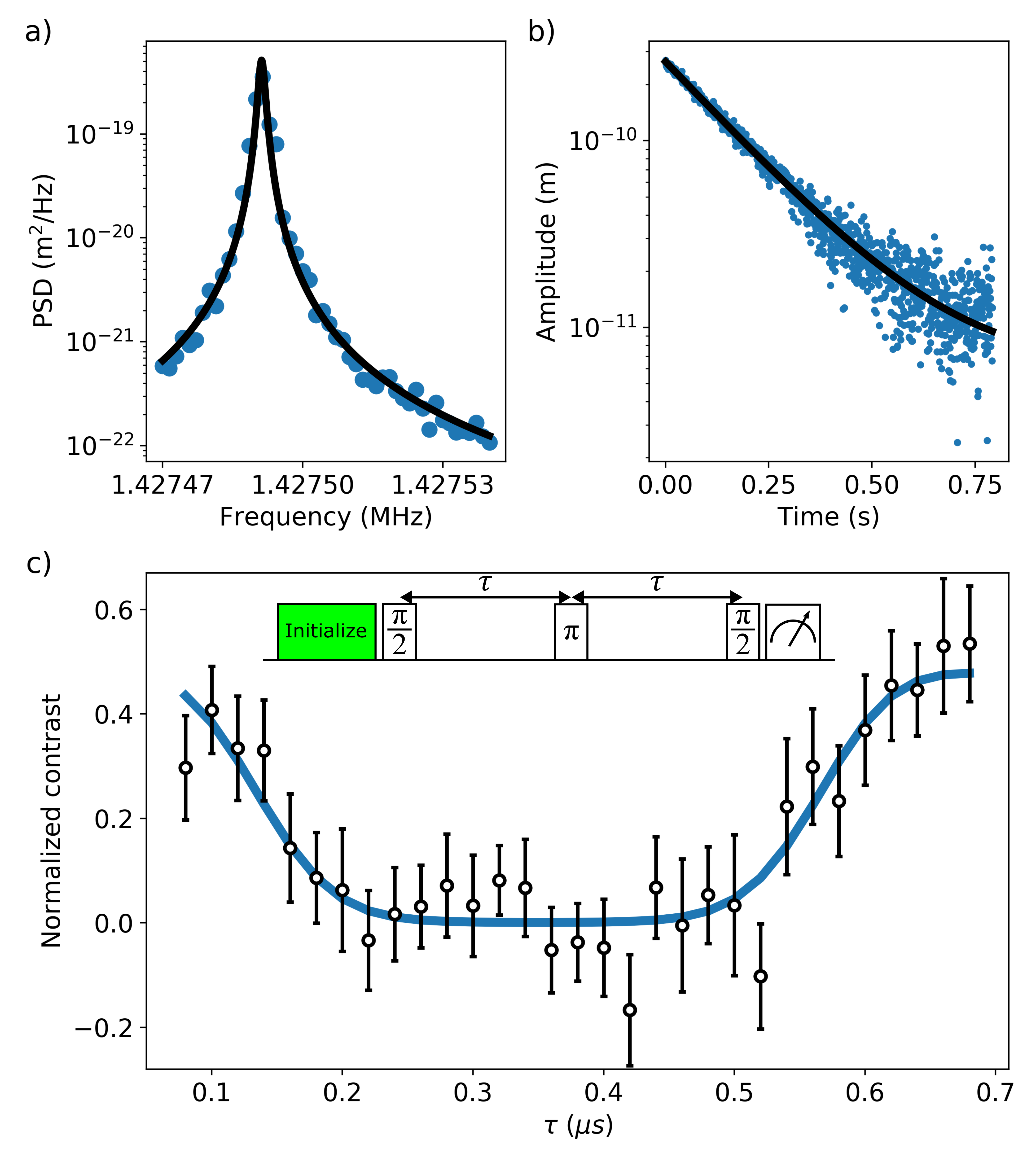}
\caption{Sensing the mechanical motion of the nanobeam. {\rm(a)} Power spectral density (PSD) of the mechanical mode, measured using an interferometer. From the Lorentzian fit (black line) we extract a resonance frequency of $\sim \SI{1.4}{\mega\hertz}$ and a linewidth of $\kappa / 2\pi = \SI{1.5(2)}{\hertz}$. {\rm(b)} After switching off an external drive from a piezoelectric chip, we measure the amplitude decay of the mechanical motion and obtain $Q = \num{8.25(6)e5}$. {\bf(c)} Sensing of the mechanical motion with the NV center. We perform a Hahn echo pulse sequence on the NV center with and without the mechanical drive, and plot their ratio (black circles) such that we can neglect the NV center's decoherence in our model for the fit ($\chi(\tau)$, blue line). We fit the signal $S(\tau, \lambda, \Delta_x)/e^{-\chi (\tau)}$ (see eq. \eqref{signal}), using fixed values of $\Delta_x$ and $\omega_r$ from interferometer measurements, to find $\lambda/2\pi = \SI{7.7(9)}{\hertz}$. The zero-point motion $z_p$ is inferred from the material densities and dimensions of the mechanical resonator.}
\label{fig:coupling} 
\end{figure}

Finally, we characterize the spin-mechanical coupling by exciting the nanobeam and characterizing its mechanical motion via independent measurements with both an interferometer and the nearby NV center. To take advantage of higher quality factors at low temperatures, we use the scanning probe setup in a helium cryostat. The interferometric measurements, shown in Fig.~\ref{fig:coupling}(a,b), reveal a quality factor of $\num{8.25(6)e5}$ (Fig.~\ref{fig:coupling}(b)), demonstrating that the quality factor remains high despite magnetic functionalization \cite{supp}. The mechanical frequency $\omega_r \sim \SI{1.4}{\mega\hertz}$ (Fig.~\ref{fig:coupling}(a)) corresponds to a period of \SI{0.7}{\micro\second}. As a result, the mechanical resonator can undergo multiple oscillations during the spin coherence time $T_{2,e}$, which is around several microseconds. The readily accessible high mechanical frequency of the nanobeam compares favorably to other spin-mechanical platforms, 
%with low mechanical frequencies in the \SI{}{\kilo\hertz} range, 
such as those featuring cantilevers, nanowires, and magnetic levitation \cite{rugar_single_2004, arcizet_single_2011, gieseler_single-spin_2020}.

A displacement of the mechanical mode by the zero-point fluctuation $z_p$ shifts the NV center spin resonance by $\lambda / 2\pi$ via the Zeeman effect, resulting in the single-phonon coupling strength $\lambda = \gamma_e z_p \nabla_z$, where $\gamma_e$ is the NV center electronic spin gyromagnetic ratio, and $\nabla_z$ is the magnetic field gradient along the NV center quantization axis. To quantify the spin-mechanical coupling strength, we excite the nanobeam with an external broadband drive and detect the field from the oscillating micromagnet with the nearby NV center. We use a Hahn echo pulse sequence, which results in frequency-dependent detection of the magnetic spin environment \cite{supp}. Sweeping the time $\tau$ between the $\pi$ pulses and assuming a Gaussian distribution of the mechanical state, the spin contrast can be approximated as
\begin{equation}\label{signal}
S(\tau, \lambda, \Delta_x) = \alpha e^{- 8 \Delta_x^2 \lambda^2 \sin^4{(\omega_r \tau /2)}/\omega_r^2 z_p^2}e^{-\chi(\tau)}
\end{equation}
where $\Delta_x$ is the root-mean-squared amplitude of motion, $\alpha$ is the spin readout contrast, and $\chi(\tau)$ describes the coherence decay from other noise sources in the diamond, such as the bath of $^{13}$C nuclear spins \cite{stanwix_coherence_2010} \cite{supp}. 

To determine $\lambda$, we first independently quantify $\Delta_x$ by integrating the interferometer signal of the mechanical response from the wideband drive, and assign $\omega_r$ to the center frequency. For the data corresponding to figure \ref{fig:coupling}(a), we find that $\Delta_x = 1.86(1)$ nm. We then fit the Hahn echo data, normalized to a baseline Hahn echo measurement to compensate for intrinsic NV decoherence $e^{-\chi(\tau)}$ (Fig. \ref{fig:coupling}(c), black dots). For the fit (blue line), $\omega_r$ and $\Delta_x$ are fixed, leaving $\lambda$ as a free parameter. We find that $\lambda/2\pi = \SI{7.7(9)}{\hertz}$ (Fig. \ref{fig:coupling}(b), corresponding to a gradient of \SI{2.4(1)e4}{\tesla/\meter}, similar to the gradients from the static field imaging of the same magnet (Fig. \ref{fig:field_scan}(b)).

\paragraph*{Discussion and Outlook.}
Our experiments demonstrate the feasibility of the proposed architecture for programmable mechanically-mediated interactions between distant spins. Specifically, we showed that the NV center's intrinsic nuclear spin memory is not degraded by movement inside a field gradient, if the proper decoupling pulse sequences are applied. The demonstrated movement distance of $\SI{1.7(2)}{\micro\meter}$ in Fig.~\ref{fig:coherence}(a-c) significantly exceeds the range of magnetic dipole-dipole interactions between spins, and is limited by the moving speed of $\SI{1}{\milli\meter/\second}$ and nuclear spin coherence time $T_{2,n} \sim 5\,\SI{}{\milli\second}$ \cite{supp}. The speed can be increased by using a nanopositioner with a higher bandwidth and minimizing residual vibrations caused by the scanning motion. By decoupling the nuclear spin from its local environment or cooling to cryogenic temperatures, $T_{2,n}$ can be extended to $>1\,\SI{}{\second}$ \cite{maurer2012room, pfender2017protecting}, which would extend the possible distance to $> 1\,\SI{}{\milli\meter}$ even with the current speed.

At the same time, achieving quantum coherent spin-mechanical coupling \cite{rabl_cooling_2010, bennett_measuring_2012, vinante_upper_2016, van_wezel_nanoscale_2012, schuetz_high-fidelity_2017, rosenfeld_efficient_2020} requires increasing the coupling strength while minimizing noise. Specifically, the onset of coherent quantum phenomena is generally marked by the spin-mechanical cooperativity $C \equiv \frac{\lambda ^2}{\Gamma \kappa n_{th}} \gtrsim 1$, which compares the coherent coupling rate $\lambda$ to the dissipation rates $\Gamma, \kappa n_{th}$ of the spin and mechanical mode respectively. While the cooperativity of our present experiment exceeds previous spin-mechanical platforms involving NV centers \cite{supp}, it remains far below the coherent coupling regime. However, significant improvements can be made. Drift of the NV-magnet distance causes significant variations of the ESR frequency at high magnetic field gradients, limiting our current gradient to $\SI{2.4e4}{\tesla/\meter}$ at a distance of \SI{1.0}{\micro \meter}. With improvements to the setup stability and the use of atomic-force microscopy (AFM) feedback \cite{jayich_afm,zhou_scanning_2017}, positioning the NV center at a reduced distance of \SI{50}{\nano \meter} from the surface of a \SI{1}{\micro\meter}-diameter micromagnet should yield gradients $\sim\SI{1.4e6}{\tesla/\meter}$, or a spin-mechanical coupling of $\lambda/2\pi \sim \SI{800}{\hertz}$. The doubly clamped nanobeam can be replaced with recent designs that utilize strain engineering and soft-clamping, which have demonstrated $Q \sim 10^9$ at MHz frequencies \cite{tsaturyan_ultracoherent_2017, ghadimi_elastic_2018, kippenberg_fractal_2021, grob_fractal_2022, bereyhi2022perimeter}. Even higher quality factors have been demonstrated or predicted, by replacing silicon nitride with crystalline materials such as silicon and diamond \cite{beccari2022strained,sementilli2022nanomechanical}. For a coupling strength of $\lambda/2\pi = \SI{800}{\hertz}$, an NV center electronic spin coherence time $T_{2,e}$ of \SI{10}{\milli\second} \cite{abobeih2018one,bar-gill_solid-state_2013}, and a quality factor of $10^9$ at \SI{4}{\kelvin}, the coherent coupling regime is possible with $C \sim 75$. Under such conditions, mechanics-mediated entanglement of electronic spins with fidelity exceeding \SI{95}{\percent} should be feasible according to the proposal in \cite{rosenfeld_efficient_2020}. Although we expect $T_{2,e}$ to improve with larger NV implantation depth, further investigation into diamond fabrication and surface termination might be required to increase $T_{2,e}$ to the \SI{10}{\milli\second} regime for NV centers in diamond nanopillars \cite{luan_decoherence_nodate, sangtawesin_origins_2019}.

Compared to previous work involving on-chip, circuit-based hybrid quantum systems \cite{rabl_quantum_2010, oconnell_quantum_2010}, a spin-mechanical architecture featuring dynamical qubit transport has the advantage of being able to generate programmable, non-local interactions, similar to reconfigurable platforms based on neutral atoms and trapped ions. The long coherence time of the nuclear spin allows multiple distant spins to be dynamically transported to interact with the same mechanical bus. While Fig.~\ref{fig:setup} only shows one mechanical resonator for clarity, the architecture can also be parallelized, with multiple mechanical resonators simultaneously mediating interactions within large arrays of spins. We also note that, unlike most other hybrid quantum systems \cite{gieseler_single-spin_2020, oconnell_quantum_2010}, both the mechanical and spin components of our platform are highly coherent even at room temperature. With a nanomagnet diameter of \SI{0.3}{\micro\meter}, an NV-magnet separation of \SI{20}{\nano \meter}, spin coherence time of $T_{2,e} = \SI{2}{\milli\second}$ \cite{herbschleb2019ultra}, and $Q=\num{1e9}$, reaching the coherent-coupling regime at room temperature appears feasible. The above considerations indicate that with realistic improvements, our platform can enable programmable interactions between distant spins, opening up a new avenue towards scalable quantum information processing with solid-state spin qubits. Finally, the present approach can be extended to realize other hybrid systems by coupling spin qubits to  quantum systems such as superconducting qubits and optical photons \cite{clerk2020hybrid,chu2020perspective,barzanjeh2022optomechanics}.

\vspace{10pt}
\paragraph*{Acknowledgements} We thank M. Markham and Element Six for providing diamond samples, R. Walsworth and M. Turner for annealing diamonds, K. Van Kirk, D. Bluvstein, A. Mulski, and E. Polzik for helpful discussions, B. Machielse for assistance with fabrication, R. Riedinger for comments on the manuscript, S. Y. F. Zhou and A. Cui for assistance with sample magnetization, and A. S. Zibrov and J. MacArthur for technical assistance. This work was supported by NSF, Center for Ultracold Atoms, and DOE Quantum Systems Accelerator Center (contract no. 7568717). E. R. acknowledges support from the NSF Graduate Research Fellowship Program. A. K. acknowledges support from the DOD through the NDSEG Fellowship Program. J. G. acknowledges support by the European Union (SEQOO, H2020-MSCA-IF-2014, No. 655369). Diamond and resonator sample fabrication were performed at the Center for Nanoscale Systems (CNS), a member of the National Nanotechnology Coordinated Infrastructure, which
is supported by the NSF under award no. ECCS -1541959. CNS is part of Harvard University. T. X. Z is an affiliate to Center for Quantum Network (CQN) established by NSF to facilitate collaboration within CQN to advance quantum research. T.M. acknowledges support from the NSF Graduate Research Fellowship Program (grant 2140743). N.A. acknowledges support from the Alexander von Humboldt Foundation and by the Federal Ministry of Education and Research (BMBF, project ``13N16297'').

\bibliography{strings} 

%apsrev4-2.bst 2019-01-14 (MD) hand-edited version of apsrev4-1.bst
%Control: key (0)
%Control: author (8) initials jnrlst
%Control: editor formatted (1) identically to author
%Control: production of article title (0) allowed
%Control: page (0) single
%Control: year (1) truncated
%Control: production of eprint (0) enabled
\begin{thebibliography}{62}%
\makeatletter
\providecommand \@ifxundefined [1]{%
 \@ifx{#1\undefined}
}%
\providecommand \@ifnum [1]{%
 \ifnum #1\expandafter \@firstoftwo
 \else \expandafter \@secondoftwo
 \fi
}%
\providecommand \@ifx [1]{%
 \ifx #1\expandafter \@firstoftwo
 \else \expandafter \@secondoftwo
 \fi
}%
\providecommand \natexlab [1]{#1}%
\providecommand \enquote  [1]{``#1''}%
\providecommand \bibnamefont  [1]{#1}%
\providecommand \bibfnamefont [1]{#1}%
\providecommand \citenamefont [1]{#1}%
\providecommand \href@noop [0]{\@secondoftwo}%
\providecommand \href [0]{\begingroup \@sanitize@url \@href}%
\providecommand \@href[1]{\@@startlink{#1}\@@href}%
\providecommand \@@href[1]{\endgroup#1\@@endlink}%
\providecommand \@sanitize@url [0]{\catcode `\\12\catcode `\$12\catcode
  `\&12\catcode `\#12\catcode `\^12\catcode `\_12\catcode `\%12\relax}%
\providecommand \@@startlink[1]{}%
\providecommand \@@endlink[0]{}%
\providecommand \url  [0]{\begingroup\@sanitize@url \@url }%
\providecommand \@url [1]{\endgroup\@href {#1}{\urlprefix }}%
\providecommand \urlprefix  [0]{URL }%
\providecommand \Eprint [0]{\href }%
\providecommand \doibase [0]{https://doi.org/}%
\providecommand \selectlanguage [0]{\@gobble}%
\providecommand \bibinfo  [0]{\@secondoftwo}%
\providecommand \bibfield  [0]{\@secondoftwo}%
\providecommand \translation [1]{[#1]}%
\providecommand \BibitemOpen [0]{}%
\providecommand \bibitemStop [0]{}%
\providecommand \bibitemNoStop [0]{.\EOS\space}%
\providecommand \EOS [0]{\spacefactor3000\relax}%
\providecommand \BibitemShut  [1]{\csname bibitem#1\endcsname}%
\let\auto@bib@innerbib\@empty
%</preamble>
\bibitem [{\citenamefont {Balasubramanian}\ \emph {et~al.}(2009)\citenamefont
  {Balasubramanian}, \citenamefont {Neumann}, \citenamefont {Twitchen},
  \citenamefont {Markham}, \citenamefont {Kolesov}, \citenamefont {Mizuochi},
  \citenamefont {Isoya}, \citenamefont {Achard}, \citenamefont {Beck},
  \citenamefont {Tissler}, \citenamefont {Jacques}, \citenamefont {Hemmer},
  \citenamefont {Jelezko},\ and\ \citenamefont
  {Wrachtrup}}]{balasubramanian2009ultralong}%
  \BibitemOpen
  \bibfield  {author} {\bibinfo {author} {\bibfnamefont {G.}~\bibnamefont
  {Balasubramanian}}, \bibinfo {author} {\bibfnamefont {P.}~\bibnamefont
  {Neumann}}, \bibinfo {author} {\bibfnamefont {D.}~\bibnamefont {Twitchen}},
  \bibinfo {author} {\bibfnamefont {M.}~\bibnamefont {Markham}}, \bibinfo
  {author} {\bibfnamefont {R.}~\bibnamefont {Kolesov}}, \bibinfo {author}
  {\bibfnamefont {N.}~\bibnamefont {Mizuochi}}, \bibinfo {author}
  {\bibfnamefont {J.}~\bibnamefont {Isoya}}, \bibinfo {author} {\bibfnamefont
  {J.}~\bibnamefont {Achard}}, \bibinfo {author} {\bibfnamefont
  {J.}~\bibnamefont {Beck}}, \bibinfo {author} {\bibfnamefont {J.}~\bibnamefont
  {Tissler}}, \bibinfo {author} {\bibfnamefont {V.}~\bibnamefont {Jacques}},
  \bibinfo {author} {\bibfnamefont {P.~R.}\ \bibnamefont {Hemmer}}, \bibinfo
  {author} {\bibfnamefont {F.}~\bibnamefont {Jelezko}},\ and\ \bibinfo {author}
  {\bibfnamefont {J.}~\bibnamefont {Wrachtrup}},\ }\bibfield  {title} {\bibinfo
  {title} {Ultralong spin coherence time in isotopically engineered diamond},\
  }\href {https://doi.org/10.1038/nmat2420} {\bibfield  {journal} {\bibinfo
  {journal} {Nat. Mat.}\ }\textbf {\bibinfo {volume} {8}},\ \bibinfo {pages}
  {383} (\bibinfo {year} {2009})}\BibitemShut {NoStop}%
\bibitem [{\citenamefont {Maurer}\ \emph {et~al.}(2012)\citenamefont {Maurer},
  \citenamefont {Kucsko}, \citenamefont {Latta}, \citenamefont {Jiang},
  \citenamefont {Yao}, \citenamefont {Bennett}, \citenamefont {Pastawski},
  \citenamefont {Hunger}, \citenamefont {Chisholm}, \citenamefont {Markham},
  \citenamefont {Twitchen}, \citenamefont {Cirac},\ and\ \citenamefont
  {Lukin}}]{maurer2012room}%
  \BibitemOpen
  \bibfield  {author} {\bibinfo {author} {\bibfnamefont {P.~C.}\ \bibnamefont
  {Maurer}}, \bibinfo {author} {\bibfnamefont {G.}~\bibnamefont {Kucsko}},
  \bibinfo {author} {\bibfnamefont {C.}~\bibnamefont {Latta}}, \bibinfo
  {author} {\bibfnamefont {L.}~\bibnamefont {Jiang}}, \bibinfo {author}
  {\bibfnamefont {N.~Y.}\ \bibnamefont {Yao}}, \bibinfo {author} {\bibfnamefont
  {S.~D.}\ \bibnamefont {Bennett}}, \bibinfo {author} {\bibfnamefont
  {F.}~\bibnamefont {Pastawski}}, \bibinfo {author} {\bibfnamefont
  {D.}~\bibnamefont {Hunger}}, \bibinfo {author} {\bibfnamefont
  {N.}~\bibnamefont {Chisholm}}, \bibinfo {author} {\bibfnamefont
  {M.}~\bibnamefont {Markham}}, \bibinfo {author} {\bibfnamefont {D.~J.}\
  \bibnamefont {Twitchen}}, \bibinfo {author} {\bibfnamefont {J.~I.}\
  \bibnamefont {Cirac}},\ and\ \bibinfo {author} {\bibfnamefont {M.~D.}\
  \bibnamefont {Lukin}},\ }\bibfield  {title} {\bibinfo {title}
  {Room-temperature quantum bit memory exceeding one second},\ }\href
  {https://doi.org/10.1126/science.1220513} {\bibfield  {journal} {\bibinfo
  {journal} {Science}\ }\textbf {\bibinfo {volume} {336}},\ \bibinfo {pages}
  {1283} (\bibinfo {year} {2012})}\BibitemShut {NoStop}%
\bibitem [{\citenamefont {Widmann}\ \emph {et~al.}(2015)\citenamefont
  {Widmann}, \citenamefont {Lee}, \citenamefont {Rendler}, \citenamefont {Son},
  \citenamefont {Fedder}, \citenamefont {Paik}, \citenamefont {Yang},
  \citenamefont {Zhao}, \citenamefont {Yang}, \citenamefont {Booker} \emph
  {et~al.}}]{widmann2015coherent}%
  \BibitemOpen
  \bibfield  {author} {\bibinfo {author} {\bibfnamefont {M.}~\bibnamefont
  {Widmann}}, \bibinfo {author} {\bibfnamefont {S.-Y.}\ \bibnamefont {Lee}},
  \bibinfo {author} {\bibfnamefont {T.}~\bibnamefont {Rendler}}, \bibinfo
  {author} {\bibfnamefont {N.~T.}\ \bibnamefont {Son}}, \bibinfo {author}
  {\bibfnamefont {H.}~\bibnamefont {Fedder}}, \bibinfo {author} {\bibfnamefont
  {S.}~\bibnamefont {Paik}}, \bibinfo {author} {\bibfnamefont {L.-P.}\
  \bibnamefont {Yang}}, \bibinfo {author} {\bibfnamefont {N.}~\bibnamefont
  {Zhao}}, \bibinfo {author} {\bibfnamefont {S.}~\bibnamefont {Yang}}, \bibinfo
  {author} {\bibfnamefont {I.}~\bibnamefont {Booker}}, \emph {et~al.},\
  }\bibfield  {title} {\bibinfo {title} {Coherent control of single spins in
  silicon carbide at room temperature},\ }\href
  {https://doi.org/10.1038/nmat4145} {\bibfield  {journal} {\bibinfo  {journal}
  {Nat. Mat.}\ }\textbf {\bibinfo {volume} {14}},\ \bibinfo {pages} {164}
  (\bibinfo {year} {2015})}\BibitemShut {NoStop}%
\bibitem [{\citenamefont {Anderson}\ \emph {et~al.}(2022)\citenamefont
  {Anderson}, \citenamefont {Glen}, \citenamefont {Zeledon}, \citenamefont
  {Bourassa}, \citenamefont {Jin}, \citenamefont {Zhu}, \citenamefont
  {Vorwerk}, \citenamefont {Crook}, \citenamefont {Abe}, \citenamefont
  {Ul-Hassan} \emph {et~al.}}]{anderson2022five}%
  \BibitemOpen
  \bibfield  {author} {\bibinfo {author} {\bibfnamefont {C.~P.}\ \bibnamefont
  {Anderson}}, \bibinfo {author} {\bibfnamefont {E.~O.}\ \bibnamefont {Glen}},
  \bibinfo {author} {\bibfnamefont {C.}~\bibnamefont {Zeledon}}, \bibinfo
  {author} {\bibfnamefont {A.}~\bibnamefont {Bourassa}}, \bibinfo {author}
  {\bibfnamefont {Y.}~\bibnamefont {Jin}}, \bibinfo {author} {\bibfnamefont
  {Y.}~\bibnamefont {Zhu}}, \bibinfo {author} {\bibfnamefont {C.}~\bibnamefont
  {Vorwerk}}, \bibinfo {author} {\bibfnamefont {A.~L.}\ \bibnamefont {Crook}},
  \bibinfo {author} {\bibfnamefont {H.}~\bibnamefont {Abe}}, \bibinfo {author}
  {\bibfnamefont {J.}~\bibnamefont {Ul-Hassan}}, \emph {et~al.},\ }\bibfield
  {title} {\bibinfo {title} {Five-second coherence of a single spin with
  single-shot readout in silicon carbide},\ }\href
  {https://doi.org/10.1126/sciadv.abm5912} {\bibfield  {journal} {\bibinfo
  {journal} {Sci. Adv.}\ }\textbf {\bibinfo {volume} {8}},\ \bibinfo {pages}
  {eabm5912} (\bibinfo {year} {2022})}\BibitemShut {NoStop}%
\bibitem [{\citenamefont {Stas}\ \emph {et~al.}(2022)\citenamefont {Stas},
  \citenamefont {Huan}, \citenamefont {Machielse}, \citenamefont {Knall},
  \citenamefont {Suleymanzade}, \citenamefont {Pingault}, \citenamefont
  {Sutula}, \citenamefont {Ding}, \citenamefont {Knaut}, \citenamefont
  {Assumpcao} \emph {et~al.}}]{stas2022robust}%
  \BibitemOpen
  \bibfield  {author} {\bibinfo {author} {\bibfnamefont {P.-J.}\ \bibnamefont
  {Stas}}, \bibinfo {author} {\bibfnamefont {Y.~Q.}\ \bibnamefont {Huan}},
  \bibinfo {author} {\bibfnamefont {B.}~\bibnamefont {Machielse}}, \bibinfo
  {author} {\bibfnamefont {E.~N.}\ \bibnamefont {Knall}}, \bibinfo {author}
  {\bibfnamefont {A.}~\bibnamefont {Suleymanzade}}, \bibinfo {author}
  {\bibfnamefont {B.}~\bibnamefont {Pingault}}, \bibinfo {author}
  {\bibfnamefont {M.}~\bibnamefont {Sutula}}, \bibinfo {author} {\bibfnamefont
  {S.~W.}\ \bibnamefont {Ding}}, \bibinfo {author} {\bibfnamefont {C.~M.}\
  \bibnamefont {Knaut}}, \bibinfo {author} {\bibfnamefont {D.~R.}\ \bibnamefont
  {Assumpcao}}, \emph {et~al.},\ }\bibfield  {title} {\bibinfo {title} {Robust
  multi-qubit quantum network node with integrated error detection},\ }\href
  {https://doi.org/10.1126/science.add9771} {\bibfield  {journal} {\bibinfo
  {journal} {Science}\ }\textbf {\bibinfo {volume} {378}},\ \bibinfo {pages}
  {557} (\bibinfo {year} {2022})}\BibitemShut {NoStop}%
\bibitem [{\citenamefont {Dutt}\ \emph {et~al.}(2007)\citenamefont {Dutt},
  \citenamefont {Childress}, \citenamefont {Jiang}, \citenamefont {Togan},
  \citenamefont {Maze}, \citenamefont {Jelezko}, \citenamefont {Zibrov},
  \citenamefont {Hemmer},\ and\ \citenamefont {Lukin}}]{dutt2007quantum}%
  \BibitemOpen
  \bibfield  {author} {\bibinfo {author} {\bibfnamefont {M.~V.~G.}\
  \bibnamefont {Dutt}}, \bibinfo {author} {\bibfnamefont {L.}~\bibnamefont
  {Childress}}, \bibinfo {author} {\bibfnamefont {L.}~\bibnamefont {Jiang}},
  \bibinfo {author} {\bibfnamefont {E.}~\bibnamefont {Togan}}, \bibinfo
  {author} {\bibfnamefont {J.}~\bibnamefont {Maze}}, \bibinfo {author}
  {\bibfnamefont {F.}~\bibnamefont {Jelezko}}, \bibinfo {author} {\bibfnamefont
  {A.}~\bibnamefont {Zibrov}}, \bibinfo {author} {\bibfnamefont
  {P.}~\bibnamefont {Hemmer}},\ and\ \bibinfo {author} {\bibfnamefont
  {M.}~\bibnamefont {Lukin}},\ }\bibfield  {title} {\bibinfo {title} {Quantum
  register based on individual electronic and nuclear spin qubits in diamond},\
  }\href {https://doi.org/10.1126/science.1139831} {\bibfield  {journal}
  {\bibinfo  {journal} {Science}\ }\textbf {\bibinfo {volume} {316}},\ \bibinfo
  {pages} {1312} (\bibinfo {year} {2007})}\BibitemShut {NoStop}%
\bibitem [{\citenamefont {Abobeih}\ \emph {et~al.}(2018)\citenamefont
  {Abobeih}, \citenamefont {Cramer}, \citenamefont {Bakker}, \citenamefont
  {Kalb}, \citenamefont {Markham}, \citenamefont {Twitchen},\ and\
  \citenamefont {Taminiau}}]{abobeih2018one}%
  \BibitemOpen
  \bibfield  {author} {\bibinfo {author} {\bibfnamefont {M.~H.}\ \bibnamefont
  {Abobeih}}, \bibinfo {author} {\bibfnamefont {J.}~\bibnamefont {Cramer}},
  \bibinfo {author} {\bibfnamefont {M.~A.}\ \bibnamefont {Bakker}}, \bibinfo
  {author} {\bibfnamefont {N.}~\bibnamefont {Kalb}}, \bibinfo {author}
  {\bibfnamefont {M.}~\bibnamefont {Markham}}, \bibinfo {author} {\bibfnamefont
  {D.~J.}\ \bibnamefont {Twitchen}},\ and\ \bibinfo {author} {\bibfnamefont
  {T.~H.}\ \bibnamefont {Taminiau}},\ }\bibfield  {title} {\bibinfo {title}
  {One-second coherence for a single electron spin coupled to a multi-qubit
  nuclear-spin environment},\ }\href
  {https://doi.org/10.1038/s41467-018-04916-z} {\bibfield  {journal} {\bibinfo
  {journal} {Nat. Commun.}\ }\textbf {\bibinfo {volume} {9}},\ \bibinfo {pages}
  {2552} (\bibinfo {year} {2018})}\BibitemShut {NoStop}%
\bibitem [{\citenamefont {Bradley}\ \emph {et~al.}(2019)\citenamefont
  {Bradley}, \citenamefont {Randall}, \citenamefont {Abobeih}, \citenamefont
  {Berrevoets}, \citenamefont {Degen}, \citenamefont {Bakker}, \citenamefont
  {Markham}, \citenamefont {Twitchen},\ and\ \citenamefont
  {Taminiau}}]{bradley2019ten}%
  \BibitemOpen
  \bibfield  {author} {\bibinfo {author} {\bibfnamefont {C.~E.}\ \bibnamefont
  {Bradley}}, \bibinfo {author} {\bibfnamefont {J.}~\bibnamefont {Randall}},
  \bibinfo {author} {\bibfnamefont {M.~H.}\ \bibnamefont {Abobeih}}, \bibinfo
  {author} {\bibfnamefont {R.~C.}\ \bibnamefont {Berrevoets}}, \bibinfo
  {author} {\bibfnamefont {M.~J.}\ \bibnamefont {Degen}}, \bibinfo {author}
  {\bibfnamefont {M.~A.}\ \bibnamefont {Bakker}}, \bibinfo {author}
  {\bibfnamefont {M.}~\bibnamefont {Markham}}, \bibinfo {author} {\bibfnamefont
  {D.~J.}\ \bibnamefont {Twitchen}},\ and\ \bibinfo {author} {\bibfnamefont
  {T.~H.}\ \bibnamefont {Taminiau}},\ }\bibfield  {title} {\bibinfo {title} {A
  ten-qubit solid-state spin register with quantum memory up to one minute},\
  }\href {https://link.aps.org/doi/10.1103/PhysRevX.9.031045} {\bibfield
  {journal} {\bibinfo  {journal} {Phys. Rev. X}\ }\textbf {\bibinfo {volume}
  {9}},\ \bibinfo {pages} {031045} (\bibinfo {year} {2019})}\BibitemShut
  {NoStop}%
\bibitem [{\citenamefont {Bhaskar}\ \emph {et~al.}(2020)\citenamefont
  {Bhaskar}, \citenamefont {Riedinger}, \citenamefont {Machielse},
  \citenamefont {Levonian}, \citenamefont {Nguyen}, \citenamefont {Knall},
  \citenamefont {Park}, \citenamefont {Englund}, \citenamefont {Lon{\v{c}}ar},
  \citenamefont {Sukachev} \emph {et~al.}}]{bhaskar2020experimental}%
  \BibitemOpen
  \bibfield  {author} {\bibinfo {author} {\bibfnamefont {M.~K.}\ \bibnamefont
  {Bhaskar}}, \bibinfo {author} {\bibfnamefont {R.}~\bibnamefont {Riedinger}},
  \bibinfo {author} {\bibfnamefont {B.}~\bibnamefont {Machielse}}, \bibinfo
  {author} {\bibfnamefont {D.~S.}\ \bibnamefont {Levonian}}, \bibinfo {author}
  {\bibfnamefont {C.~T.}\ \bibnamefont {Nguyen}}, \bibinfo {author}
  {\bibfnamefont {E.~N.}\ \bibnamefont {Knall}}, \bibinfo {author}
  {\bibfnamefont {H.}~\bibnamefont {Park}}, \bibinfo {author} {\bibfnamefont
  {D.}~\bibnamefont {Englund}}, \bibinfo {author} {\bibfnamefont
  {M.}~\bibnamefont {Lon{\v{c}}ar}}, \bibinfo {author} {\bibfnamefont {D.~D.}\
  \bibnamefont {Sukachev}}, \emph {et~al.},\ }\bibfield  {title} {\bibinfo
  {title} {Experimental demonstration of memory-enhanced quantum
  communication},\ }\href {https://doi.org/10.1038/s41586-020-2103-5}
  {\bibfield  {journal} {\bibinfo  {journal} {Nature}\ }\textbf {\bibinfo
  {volume} {580}},\ \bibinfo {pages} {60} (\bibinfo {year} {2020})}\BibitemShut
  {NoStop}%
\bibitem [{\citenamefont {Hermans}\ \emph {et~al.}(2022)\citenamefont
  {Hermans}, \citenamefont {Pompili}, \citenamefont {Beukers}, \citenamefont
  {Baier}, \citenamefont {Borregaard},\ and\ \citenamefont
  {Hanson}}]{hermans2022qubit}%
  \BibitemOpen
  \bibfield  {author} {\bibinfo {author} {\bibfnamefont {S.~L.~N.}\
  \bibnamefont {Hermans}}, \bibinfo {author} {\bibfnamefont {M.}~\bibnamefont
  {Pompili}}, \bibinfo {author} {\bibfnamefont {H.~K.~C.}\ \bibnamefont
  {Beukers}}, \bibinfo {author} {\bibfnamefont {S.}~\bibnamefont {Baier}},
  \bibinfo {author} {\bibfnamefont {J.}~\bibnamefont {Borregaard}},\ and\
  \bibinfo {author} {\bibfnamefont {R.}~\bibnamefont {Hanson}},\ }\bibfield
  {title} {\bibinfo {title} {Qubit teleportation between non-neighbouring nodes
  in a quantum network},\ }\href {https://doi.org/10.1038/s41586-022-04697-y}
  {\bibfield  {journal} {\bibinfo  {journal} {Nature}\ }\textbf {\bibinfo
  {volume} {605}},\ \bibinfo {pages} {663} (\bibinfo {year}
  {2022})}\BibitemShut {NoStop}%
\bibitem [{\citenamefont {Rabl}\ \emph {et~al.}(2010)\citenamefont {Rabl},
  \citenamefont {Kolkowitz}, \citenamefont {Koppens}, \citenamefont {Harris},
  \citenamefont {Zoller},\ and\ \citenamefont {Lukin}}]{rabl_quantum_2010}%
  \BibitemOpen
  \bibfield  {author} {\bibinfo {author} {\bibfnamefont {P.}~\bibnamefont
  {Rabl}}, \bibinfo {author} {\bibfnamefont {S.~J.}\ \bibnamefont {Kolkowitz}},
  \bibinfo {author} {\bibfnamefont {F.~H.~L.}\ \bibnamefont {Koppens}},
  \bibinfo {author} {\bibfnamefont {J.~G.~E.}\ \bibnamefont {Harris}}, \bibinfo
  {author} {\bibfnamefont {P.}~\bibnamefont {Zoller}},\ and\ \bibinfo {author}
  {\bibfnamefont {M.~D.}\ \bibnamefont {Lukin}},\ }\bibfield  {title} {\bibinfo
  {title} {A quantum spin transducer based on nanoelectromechanical resonator
  arrays},\ }\href {https://doi.org/10.1038/nphys1679} {\bibfield  {journal}
  {\bibinfo  {journal} {Nat. Phys.}\ }\textbf {\bibinfo {volume} {6}},\
  \bibinfo {pages} {602} (\bibinfo {year} {2010})}\BibitemShut {NoStop}%
\bibitem [{\citenamefont {Kuzyk}\ and\ \citenamefont
  {Wang}(2018)}]{kuzyk2018scaling}%
  \BibitemOpen
  \bibfield  {author} {\bibinfo {author} {\bibfnamefont {M.~C.}\ \bibnamefont
  {Kuzyk}}\ and\ \bibinfo {author} {\bibfnamefont {H.}~\bibnamefont {Wang}},\
  }\bibfield  {title} {\bibinfo {title} {Scaling phononic quantum networks of
  solid-state spins with closed mechanical subsystems},\ }\href
  {https://link.aps.org/doi/10.1103/PhysRevX.8.041027} {\bibfield  {journal}
  {\bibinfo  {journal} {Phys. Rev. X}\ }\textbf {\bibinfo {volume} {8}},\
  \bibinfo {pages} {041027} (\bibinfo {year} {2018})}\BibitemShut {NoStop}%
\bibitem [{\citenamefont {Arcizet}\ \emph {et~al.}(2011)\citenamefont
  {Arcizet}, \citenamefont {Jacques}, \citenamefont {Siria}, \citenamefont
  {Poncharal}, \citenamefont {Vincent},\ and\ \citenamefont
  {Seidelin}}]{arcizet_single_2011}%
  \BibitemOpen
  \bibfield  {author} {\bibinfo {author} {\bibfnamefont {O.}~\bibnamefont
  {Arcizet}}, \bibinfo {author} {\bibfnamefont {V.}~\bibnamefont {Jacques}},
  \bibinfo {author} {\bibfnamefont {A.}~\bibnamefont {Siria}}, \bibinfo
  {author} {\bibfnamefont {P.}~\bibnamefont {Poncharal}}, \bibinfo {author}
  {\bibfnamefont {P.}~\bibnamefont {Vincent}},\ and\ \bibinfo {author}
  {\bibfnamefont {S.}~\bibnamefont {Seidelin}},\ }\bibfield  {title} {\bibinfo
  {title} {A single nitrogen-vacancy defect coupled to a nanomechanical
  oscillator},\ }\href {https://doi.org/10.1038/nphys2070} {\bibfield
  {journal} {\bibinfo  {journal} {Nat. Phys.}\ }\textbf {\bibinfo {volume}
  {7}},\ \bibinfo {pages} {879} (\bibinfo {year} {2011})}\BibitemShut {NoStop}%
\bibitem [{\citenamefont {Bennett}\ \emph {et~al.}(2012)\citenamefont
  {Bennett}, \citenamefont {Kolkowitz}, \citenamefont {Unterreithmeier},
  \citenamefont {Rabl}, \citenamefont {Jayich}, \citenamefont {Harris},\ and\
  \citenamefont {Lukin}}]{bennett_measuring_2012}%
  \BibitemOpen
  \bibfield  {author} {\bibinfo {author} {\bibfnamefont {S.~D.}\ \bibnamefont
  {Bennett}}, \bibinfo {author} {\bibfnamefont {S.}~\bibnamefont {Kolkowitz}},
  \bibinfo {author} {\bibfnamefont {Q.~P.}\ \bibnamefont {Unterreithmeier}},
  \bibinfo {author} {\bibfnamefont {P.}~\bibnamefont {Rabl}}, \bibinfo {author}
  {\bibfnamefont {A.~C.~B.}\ \bibnamefont {Jayich}}, \bibinfo {author}
  {\bibfnamefont {J.~G.~E.}\ \bibnamefont {Harris}},\ and\ \bibinfo {author}
  {\bibfnamefont {M.~D.}\ \bibnamefont {Lukin}},\ }\bibfield  {title} {\bibinfo
  {title} {Measuring mechanical motion with a single spin},\ }\href
  {doi.org/10.1088/1367-2630/14/12/125004} {\bibfield  {journal} {\bibinfo
  {journal} {New J. Phys.}\ }\textbf {\bibinfo {volume} {14}},\ \bibinfo
  {pages} {125004} (\bibinfo {year} {2012})}\BibitemShut {NoStop}%
\bibitem [{\citenamefont {Kolkowitz}\ \emph {et~al.}(2012)\citenamefont
  {Kolkowitz}, \citenamefont {Jayich}, \citenamefont {Unterreithmeier},
  \citenamefont {Bennett}, \citenamefont {Rabl}, \citenamefont {Harris},\ and\
  \citenamefont {Lukin}}]{kolkowitz_coherent_2012}%
  \BibitemOpen
  \bibfield  {author} {\bibinfo {author} {\bibfnamefont {S.}~\bibnamefont
  {Kolkowitz}}, \bibinfo {author} {\bibfnamefont {A.~C.~B.}\ \bibnamefont
  {Jayich}}, \bibinfo {author} {\bibfnamefont {Q.~P.}\ \bibnamefont
  {Unterreithmeier}}, \bibinfo {author} {\bibfnamefont {S.~D.}\ \bibnamefont
  {Bennett}}, \bibinfo {author} {\bibfnamefont {P.}~\bibnamefont {Rabl}},
  \bibinfo {author} {\bibfnamefont {J.~G.~E.}\ \bibnamefont {Harris}},\ and\
  \bibinfo {author} {\bibfnamefont {M.~D.}\ \bibnamefont {Lukin}},\ }\bibfield
  {title} {\bibinfo {title} {Coherent sensing of a mechanical resonator with a
  single-spin qubit},\ }\href {https://doi.org/10.1126/science.1216821}
  {\bibfield  {journal} {\bibinfo  {journal} {Science}\ }\textbf {\bibinfo
  {volume} {335}},\ \bibinfo {pages} {1603} (\bibinfo {year}
  {2012})}\BibitemShut {NoStop}%
\bibitem [{\citenamefont {Nichol}\ \emph {et~al.}(2012)\citenamefont {Nichol},
  \citenamefont {Hemesath}, \citenamefont {Lauhon},\ and\ \citenamefont
  {Budakian}}]{nichol_nanomechanical_2012}%
  \BibitemOpen
  \bibfield  {author} {\bibinfo {author} {\bibfnamefont {J.~M.}\ \bibnamefont
  {Nichol}}, \bibinfo {author} {\bibfnamefont {E.~R.}\ \bibnamefont
  {Hemesath}}, \bibinfo {author} {\bibfnamefont {L.~J.}\ \bibnamefont
  {Lauhon}},\ and\ \bibinfo {author} {\bibfnamefont {R.}~\bibnamefont
  {Budakian}},\ }\bibfield  {title} {\bibinfo {title} {Nanomechanical detection
  of nuclear magnetic resonance using a silicon nanowire oscillator},\ }\href
  {https://doi.org/10.1103/PhysRevB.85.054414} {\bibfield  {journal} {\bibinfo
  {journal} {Phys. Rev. B}\ }\textbf {\bibinfo {volume} {85}},\ \bibinfo
  {pages} {054414} (\bibinfo {year} {2012})}\BibitemShut {NoStop}%
\bibitem [{\citenamefont {Rugar}\ \emph {et~al.}(2004)\citenamefont {Rugar},
  \citenamefont {Budakian}, \citenamefont {Mamin},\ and\ \citenamefont
  {Chui}}]{rugar_single_2004}%
  \BibitemOpen
  \bibfield  {author} {\bibinfo {author} {\bibfnamefont {D.}~\bibnamefont
  {Rugar}}, \bibinfo {author} {\bibfnamefont {R.}~\bibnamefont {Budakian}},
  \bibinfo {author} {\bibfnamefont {H.~J.}\ \bibnamefont {Mamin}},\ and\
  \bibinfo {author} {\bibfnamefont {B.~W.}\ \bibnamefont {Chui}},\ }\bibfield
  {title} {\bibinfo {title} {Single spin detection by magnetic resonance force
  microscopy},\ }\href {https://doi.org/10.1038/nature02658} {\bibfield
  {journal} {\bibinfo  {journal} {Nature}\ }\textbf {\bibinfo {volume} {430}},\
  \bibinfo {pages} {329} (\bibinfo {year} {2004})}\BibitemShut {NoStop}%
\bibitem [{\citenamefont {Teissier}\ \emph {et~al.}(2014)\citenamefont
  {Teissier}, \citenamefont {Barfuss}, \citenamefont {Appel}, \citenamefont
  {Neu},\ and\ \citenamefont {Maletinsky}}]{teissier_strain_2014}%
  \BibitemOpen
  \bibfield  {author} {\bibinfo {author} {\bibfnamefont {J.}~\bibnamefont
  {Teissier}}, \bibinfo {author} {\bibfnamefont {A.}~\bibnamefont {Barfuss}},
  \bibinfo {author} {\bibfnamefont {P.}~\bibnamefont {Appel}}, \bibinfo
  {author} {\bibfnamefont {E.}~\bibnamefont {Neu}},\ and\ \bibinfo {author}
  {\bibfnamefont {P.}~\bibnamefont {Maletinsky}},\ }\bibfield  {title}
  {\bibinfo {title} {Strain coupling of a nitrogen-vacancy center spin to a
  diamond mechanical oscillator},\ }\href
  {https://link.aps.org/doi/10.1103/PhysRevLett.113.020503} {\bibfield
  {journal} {\bibinfo  {journal} {Phys. Rev. Lett.}\ }\textbf {\bibinfo
  {volume} {113}},\ \bibinfo {pages} {020503} (\bibinfo {year}
  {2014})}\BibitemShut {NoStop}%
\bibitem [{\citenamefont {Pigeau}\ \emph {et~al.}(2015)\citenamefont {Pigeau},
  \citenamefont {Rohr}, \citenamefont {Mercier~de L{\'e}pinay}, \citenamefont
  {Gloppe}, \citenamefont {Jacques},\ and\ \citenamefont
  {Arcizet}}]{pigeau_observation_2015}%
  \BibitemOpen
  \bibfield  {author} {\bibinfo {author} {\bibfnamefont {B.}~\bibnamefont
  {Pigeau}}, \bibinfo {author} {\bibfnamefont {S.}~\bibnamefont {Rohr}},
  \bibinfo {author} {\bibfnamefont {L.}~\bibnamefont {Mercier~de L{\'e}pinay}},
  \bibinfo {author} {\bibfnamefont {A.}~\bibnamefont {Gloppe}}, \bibinfo
  {author} {\bibfnamefont {V.}~\bibnamefont {Jacques}},\ and\ \bibinfo {author}
  {\bibfnamefont {O.}~\bibnamefont {Arcizet}},\ }\bibfield  {title} {\bibinfo
  {title} {Observation of a phononic mollow triplet in a multimode hybrid
  spin-nanomechanical system},\ }\href {https://doi.org/10.1038/ncomms9603}
  {\bibfield  {journal} {\bibinfo  {journal} {Nat. Commun.}\ }\textbf {\bibinfo
  {volume} {6}},\ \bibinfo {pages} {8603} (\bibinfo {year} {2015})}\BibitemShut
  {NoStop}%
\bibitem [{\citenamefont {Lee}\ \emph {et~al.}(2016)\citenamefont {Lee},
  \citenamefont {Lee}, \citenamefont {Ovartchaiyapong}, \citenamefont
  {Minguzzi}, \citenamefont {Maze},\ and\ \citenamefont
  {Jayich}}]{lee_strain_2016}%
  \BibitemOpen
  \bibfield  {author} {\bibinfo {author} {\bibfnamefont {K.~W.}\ \bibnamefont
  {Lee}}, \bibinfo {author} {\bibfnamefont {D.}~\bibnamefont {Lee}}, \bibinfo
  {author} {\bibfnamefont {P.}~\bibnamefont {Ovartchaiyapong}}, \bibinfo
  {author} {\bibfnamefont {J.}~\bibnamefont {Minguzzi}}, \bibinfo {author}
  {\bibfnamefont {J.~R.}\ \bibnamefont {Maze}},\ and\ \bibinfo {author}
  {\bibfnamefont {A.~C.~B.}\ \bibnamefont {Jayich}},\ }\bibfield  {title}
  {\bibinfo {title} {Strain coupling of a mechanical resonator to a single
  quantum emitter in diamond},\ }\href
  {https://doi.org/10.1103/PhysRevApplied.6.034005} {\bibfield  {journal}
  {\bibinfo  {journal} {Phys. Rev. Appl.}\ }\textbf {\bibinfo {volume} {6}},\
  \bibinfo {pages} {034005} (\bibinfo {year} {2016})}\BibitemShut {NoStop}%
\bibitem [{\citenamefont {Meesala}\ \emph {et~al.}(2016)\citenamefont
  {Meesala}, \citenamefont {Sohn}, \citenamefont {Atikian}, \citenamefont
  {Kim}, \citenamefont {Burek}, \citenamefont {Choy},\ and\ \citenamefont
  {Lončar}}]{meesala_enhanced_2016}%
  \BibitemOpen
  \bibfield  {author} {\bibinfo {author} {\bibfnamefont {S.}~\bibnamefont
  {Meesala}}, \bibinfo {author} {\bibfnamefont {Y.-I.}\ \bibnamefont {Sohn}},
  \bibinfo {author} {\bibfnamefont {H.~A.}\ \bibnamefont {Atikian}}, \bibinfo
  {author} {\bibfnamefont {S.}~\bibnamefont {Kim}}, \bibinfo {author}
  {\bibfnamefont {M.~J.}\ \bibnamefont {Burek}}, \bibinfo {author}
  {\bibfnamefont {J.~T.}\ \bibnamefont {Choy}},\ and\ \bibinfo {author}
  {\bibfnamefont {M.}~\bibnamefont {Lončar}},\ }\bibfield  {title} {\bibinfo
  {title} {Enhanced strain coupling of nitrogen-vacancy spins to nanoscale
  diamond cantilevers},\ }\href
  {https://doi.org/10.1103/PhysRevApplied.5.034010} {\bibfield  {journal}
  {\bibinfo  {journal} {Phys. Rev. Appl.}\ }\textbf {\bibinfo {volume} {5}},\
  \bibinfo {pages} {034010} (\bibinfo {year} {2016})}\BibitemShut {NoStop}%
\bibitem [{\citenamefont {Oeckinghaus}\ \emph {et~al.}(2020)\citenamefont
  {Oeckinghaus}, \citenamefont {Momenzadeh}, \citenamefont {Scheiger},
  \citenamefont {Shalomayeva}, \citenamefont {Finkler}, \citenamefont {Dasari},
  \citenamefont {Stöhr},\ and\ \citenamefont
  {Wrachtrup}}]{wrachtrup_cantilever_2020}%
  \BibitemOpen
  \bibfield  {author} {\bibinfo {author} {\bibfnamefont {T.}~\bibnamefont
  {Oeckinghaus}}, \bibinfo {author} {\bibfnamefont {S.~A.}\ \bibnamefont
  {Momenzadeh}}, \bibinfo {author} {\bibfnamefont {P.}~\bibnamefont
  {Scheiger}}, \bibinfo {author} {\bibfnamefont {T.}~\bibnamefont
  {Shalomayeva}}, \bibinfo {author} {\bibfnamefont {A.}~\bibnamefont
  {Finkler}}, \bibinfo {author} {\bibfnamefont {D.}~\bibnamefont {Dasari}},
  \bibinfo {author} {\bibfnamefont {R.}~\bibnamefont {Stöhr}},\ and\ \bibinfo
  {author} {\bibfnamefont {J.}~\bibnamefont {Wrachtrup}},\ }\bibfield  {title}
  {\bibinfo {title} {Spin–phonon interfaces in coupled nanomechanical
  cantilevers},\ }\href {https://doi.org/10.1021/acs.nanolett.9b04198}
  {\bibfield  {journal} {\bibinfo  {journal} {Nano Lett.}\ }\textbf {\bibinfo
  {volume} {20}},\ \bibinfo {pages} {463} (\bibinfo {year} {2020})}\BibitemShut
  {NoStop}%
\bibitem [{\citenamefont {Gieseler}\ \emph {et~al.}(2020)\citenamefont
  {Gieseler}, \citenamefont {Kabcenell}, \citenamefont {Rosenfeld},
  \citenamefont {Schaefer}, \citenamefont {Safira}, \citenamefont {Schuetz},
  \citenamefont {Gonzalez-Ballestero}, \citenamefont {Rusconi}, \citenamefont
  {Romero-Isart},\ and\ \citenamefont {Lukin}}]{gieseler_single-spin_2020}%
  \BibitemOpen
  \bibfield  {author} {\bibinfo {author} {\bibfnamefont {J.}~\bibnamefont
  {Gieseler}}, \bibinfo {author} {\bibfnamefont {A.}~\bibnamefont {Kabcenell}},
  \bibinfo {author} {\bibfnamefont {E.}~\bibnamefont {Rosenfeld}}, \bibinfo
  {author} {\bibfnamefont {J.~D.}\ \bibnamefont {Schaefer}}, \bibinfo {author}
  {\bibfnamefont {A.}~\bibnamefont {Safira}}, \bibinfo {author} {\bibfnamefont
  {M.~J.~A.}\ \bibnamefont {Schuetz}}, \bibinfo {author} {\bibfnamefont
  {C.}~\bibnamefont {Gonzalez-Ballestero}}, \bibinfo {author} {\bibfnamefont
  {C.~C.}\ \bibnamefont {Rusconi}}, \bibinfo {author} {\bibfnamefont
  {O.}~\bibnamefont {Romero-Isart}},\ and\ \bibinfo {author} {\bibfnamefont
  {M.~D.}\ \bibnamefont {Lukin}},\ }\bibfield  {title} {\bibinfo {title}
  {Single-spin magnetomechanics with levitated micromagnets},\ }\href
  {https://link.aps.org/doi/10.1103/PhysRevLett.124.163604} {\bibfield
  {journal} {\bibinfo  {journal} {Phys. Rev. Lett.}\ }\textbf {\bibinfo
  {volume} {124}},\ \bibinfo {pages} {163604} (\bibinfo {year}
  {2020})}\BibitemShut {NoStop}%
\bibitem [{\citenamefont {Maity}\ \emph {et~al.}(2020)\citenamefont {Maity},
  \citenamefont {Shao}, \citenamefont {Bogdanović}, \citenamefont {Meesala},
  \citenamefont {Sohn}, \citenamefont {Sinclair}, \citenamefont {Pingault},
  \citenamefont {Chalupnik}, \citenamefont {Chia}, \citenamefont {Zheng},
  \citenamefont {Lai},\ and\ \citenamefont {Lončar}}]{maity_coherent_2020}%
  \BibitemOpen
  \bibfield  {author} {\bibinfo {author} {\bibfnamefont {S.}~\bibnamefont
  {Maity}}, \bibinfo {author} {\bibfnamefont {L.}~\bibnamefont {Shao}},
  \bibinfo {author} {\bibfnamefont {S.}~\bibnamefont {Bogdanović}}, \bibinfo
  {author} {\bibfnamefont {S.}~\bibnamefont {Meesala}}, \bibinfo {author}
  {\bibfnamefont {Y.-I.}\ \bibnamefont {Sohn}}, \bibinfo {author}
  {\bibfnamefont {N.}~\bibnamefont {Sinclair}}, \bibinfo {author}
  {\bibfnamefont {B.}~\bibnamefont {Pingault}}, \bibinfo {author}
  {\bibfnamefont {M.}~\bibnamefont {Chalupnik}}, \bibinfo {author}
  {\bibfnamefont {C.}~\bibnamefont {Chia}}, \bibinfo {author} {\bibfnamefont
  {L.}~\bibnamefont {Zheng}}, \bibinfo {author} {\bibfnamefont
  {K.}~\bibnamefont {Lai}},\ and\ \bibinfo {author} {\bibfnamefont
  {M.}~\bibnamefont {Lončar}},\ }\bibfield  {title} {\bibinfo {title}
  {Coherent acoustic control of a single silicon vacancy spin in diamond},\
  }\href {https://doi.org/10.1038/s41467-019-13822-x} {\bibfield  {journal}
  {\bibinfo  {journal} {Nat. Commun.}\ }\textbf {\bibinfo {volume} {11}},\
  \bibinfo {pages} {193} (\bibinfo {year} {2020})}\BibitemShut {NoStop}%
\bibitem [{\citenamefont {Chu}\ and\ \citenamefont
  {Gr{\"o}blacher}(2020)}]{chu2020perspective}%
  \BibitemOpen
  \bibfield  {author} {\bibinfo {author} {\bibfnamefont {Y.}~\bibnamefont
  {Chu}}\ and\ \bibinfo {author} {\bibfnamefont {S.}~\bibnamefont
  {Gr{\"o}blacher}},\ }\bibfield  {title} {\bibinfo {title} {A perspective on
  hybrid quantum opto-and electromechanical systems},\ }\href
  {https://doi.org/10.1063/5.0021088} {\bibfield  {journal} {\bibinfo
  {journal} {Appl. Phys. Lett.}\ }\textbf {\bibinfo {volume} {117}},\ \bibinfo
  {pages} {150503} (\bibinfo {year} {2020})}\BibitemShut {NoStop}%
\bibitem [{\citenamefont {Schuetz}\ \emph {et~al.}(2017)\citenamefont
  {Schuetz}, \citenamefont {Giedke}, \citenamefont {Vandersypen},\ and\
  \citenamefont {Cirac}}]{schuetz_high-fidelity_2017}%
  \BibitemOpen
  \bibfield  {author} {\bibinfo {author} {\bibfnamefont {M.~J.~A.}\
  \bibnamefont {Schuetz}}, \bibinfo {author} {\bibfnamefont {G.}~\bibnamefont
  {Giedke}}, \bibinfo {author} {\bibfnamefont {L.~M.~K.}\ \bibnamefont
  {Vandersypen}},\ and\ \bibinfo {author} {\bibfnamefont {J.~I.}\ \bibnamefont
  {Cirac}},\ }\bibfield  {title} {\bibinfo {title} {High-fidelity hot gates for
  generic spin-resonator systems},\ }\href
  {https://link.aps.org/doi/10.1103/PhysRevA.95.052335} {\bibfield  {journal}
  {\bibinfo  {journal} {Phys. Rev. A}\ }\textbf {\bibinfo {volume} {95}},\
  \bibinfo {pages} {052335} (\bibinfo {year} {2017})}\BibitemShut {NoStop}%
\bibitem [{\citenamefont {Rosenfeld}\ \emph {et~al.}(2021)\citenamefont
  {Rosenfeld}, \citenamefont {Riedinger}, \citenamefont {Gieseler},
  \citenamefont {Schuetz},\ and\ \citenamefont
  {Lukin}}]{rosenfeld_efficient_2020}%
  \BibitemOpen
  \bibfield  {author} {\bibinfo {author} {\bibfnamefont {E.}~\bibnamefont
  {Rosenfeld}}, \bibinfo {author} {\bibfnamefont {R.}~\bibnamefont
  {Riedinger}}, \bibinfo {author} {\bibfnamefont {J.}~\bibnamefont {Gieseler}},
  \bibinfo {author} {\bibfnamefont {M.}~\bibnamefont {Schuetz}},\ and\ \bibinfo
  {author} {\bibfnamefont {M.~D.}\ \bibnamefont {Lukin}},\ }\bibfield  {title}
  {\bibinfo {title} {Efficient entanglement of spin qubits mediated by a hot
  mechanical oscillator},\ }\href
  {https://doi.org/10.1103/PhysRevLett.126.250505} {\bibfield  {journal}
  {\bibinfo  {journal} {Phys. Rev. Lett.}\ }\textbf {\bibinfo {volume} {126}},\
  \bibinfo {pages} {250505} (\bibinfo {year} {2021})}\BibitemShut {NoStop}%
\bibitem [{\citenamefont {Rabl}\ \emph {et~al.}(2009)\citenamefont {Rabl},
  \citenamefont {Cappellaro}, \citenamefont {Dutt}, \citenamefont {Jiang},
  \citenamefont {Maze},\ and\ \citenamefont {Lukin}}]{rabl2009strong}%
  \BibitemOpen
  \bibfield  {author} {\bibinfo {author} {\bibfnamefont {P.}~\bibnamefont
  {Rabl}}, \bibinfo {author} {\bibfnamefont {P.}~\bibnamefont {Cappellaro}},
  \bibinfo {author} {\bibfnamefont {M.~V.~G.}\ \bibnamefont {Dutt}}, \bibinfo
  {author} {\bibfnamefont {L.}~\bibnamefont {Jiang}}, \bibinfo {author}
  {\bibfnamefont {J.~R.}\ \bibnamefont {Maze}},\ and\ \bibinfo {author}
  {\bibfnamefont {M.~D.}\ \bibnamefont {Lukin}},\ }\bibfield  {title} {\bibinfo
  {title} {Strong magnetic coupling between an electronic spin qubit and a
  mechanical resonator},\ }\href
  {https://link.aps.org/doi/10.1103/PhysRevB.79.041302} {\bibfield  {journal}
  {\bibinfo  {journal} {Phys. Rev. B}\ }\textbf {\bibinfo {volume} {79}},\
  \bibinfo {pages} {041302} (\bibinfo {year} {2009})}\BibitemShut {NoStop}%
\bibitem [{\citenamefont {Rabl}(2010)}]{rabl_cooling_2010}%
  \BibitemOpen
  \bibfield  {author} {\bibinfo {author} {\bibfnamefont {P.}~\bibnamefont
  {Rabl}},\ }\bibfield  {title} {\bibinfo {title} {Cooling of mechanical motion
  with a two-level system: The high-temperature regime},\ }\href
  {https://link.aps.org/doi/10.1103/PhysRevB.82.165320} {\bibfield  {journal}
  {\bibinfo  {journal} {Phys. Rev. B}\ }\textbf {\bibinfo {volume} {82}},\
  \bibinfo {pages} {165320} (\bibinfo {year} {2010})}\BibitemShut {NoStop}%
\bibitem [{\citenamefont {O’Connell}\ \emph {et~al.}(2010)\citenamefont
  {O’Connell}, \citenamefont {Hofheinz}, \citenamefont {Ansmann},
  \citenamefont {Bialczak}, \citenamefont {Lenander}, \citenamefont {Lucero},
  \citenamefont {Neeley}, \citenamefont {Sank}, \citenamefont {Wang},
  \citenamefont {Weides}, \citenamefont {Wenner}, \citenamefont {Martinis},\
  and\ \citenamefont {Cleland}}]{oconnell_quantum_2010}%
  \BibitemOpen
  \bibfield  {author} {\bibinfo {author} {\bibfnamefont {A.~D.}\ \bibnamefont
  {O’Connell}}, \bibinfo {author} {\bibfnamefont {M.}~\bibnamefont
  {Hofheinz}}, \bibinfo {author} {\bibfnamefont {M.}~\bibnamefont {Ansmann}},
  \bibinfo {author} {\bibfnamefont {R.~C.}\ \bibnamefont {Bialczak}}, \bibinfo
  {author} {\bibfnamefont {M.}~\bibnamefont {Lenander}}, \bibinfo {author}
  {\bibfnamefont {E.}~\bibnamefont {Lucero}}, \bibinfo {author} {\bibfnamefont
  {M.}~\bibnamefont {Neeley}}, \bibinfo {author} {\bibfnamefont
  {D.}~\bibnamefont {Sank}}, \bibinfo {author} {\bibfnamefont {H.}~\bibnamefont
  {Wang}}, \bibinfo {author} {\bibfnamefont {M.}~\bibnamefont {Weides}},
  \bibinfo {author} {\bibfnamefont {J.}~\bibnamefont {Wenner}}, \bibinfo
  {author} {\bibfnamefont {J.~M.}\ \bibnamefont {Martinis}},\ and\ \bibinfo
  {author} {\bibfnamefont {A.~N.}\ \bibnamefont {Cleland}},\ }\bibfield
  {title} {\bibinfo {title} {Quantum ground state and single-phonon control of
  a mechanical resonator},\ }\href {http://www.nature.com/articles/nature08967}
  {\bibfield  {journal} {\bibinfo  {journal} {Nature}\ }\textbf {\bibinfo
  {volume} {464}},\ \bibinfo {pages} {697} (\bibinfo {year}
  {2010})}\BibitemShut {NoStop}%
\bibitem [{\citenamefont {Tsaturyan}\ \emph {et~al.}(2017)\citenamefont
  {Tsaturyan}, \citenamefont {Barg}, \citenamefont {Polzik},\ and\
  \citenamefont {Schliesser}}]{tsaturyan_ultracoherent_2017}%
  \BibitemOpen
  \bibfield  {author} {\bibinfo {author} {\bibfnamefont {Y.}~\bibnamefont
  {Tsaturyan}}, \bibinfo {author} {\bibfnamefont {A.}~\bibnamefont {Barg}},
  \bibinfo {author} {\bibfnamefont {E.~S.}\ \bibnamefont {Polzik}},\ and\
  \bibinfo {author} {\bibfnamefont {A.}~\bibnamefont {Schliesser}},\ }\bibfield
   {title} {\bibinfo {title} {Ultracoherent nanomechanical resonators via soft
  clamping and dissipation dilution},\ }\href
  {https://doi.org/10.1038/nnano.2017.101} {\bibfield  {journal} {\bibinfo
  {journal} {Nat. Nanotechnol.}\ }\textbf {\bibinfo {volume} {12}},\ \bibinfo
  {pages} {776} (\bibinfo {year} {2017})}\BibitemShut {NoStop}%
\bibitem [{\citenamefont {Ghadimi}\ \emph {et~al.}(2018)\citenamefont
  {Ghadimi}, \citenamefont {Fedorov}, \citenamefont {Engelsen}, \citenamefont
  {Bereyhi}, \citenamefont {Schilling}, \citenamefont {Wilson},\ and\
  \citenamefont {Kippenberg}}]{ghadimi_elastic_2018}%
  \BibitemOpen
  \bibfield  {author} {\bibinfo {author} {\bibfnamefont {A.~H.}\ \bibnamefont
  {Ghadimi}}, \bibinfo {author} {\bibfnamefont {S.~A.}\ \bibnamefont
  {Fedorov}}, \bibinfo {author} {\bibfnamefont {N.~J.}\ \bibnamefont
  {Engelsen}}, \bibinfo {author} {\bibfnamefont {M.~J.}\ \bibnamefont
  {Bereyhi}}, \bibinfo {author} {\bibfnamefont {R.}~\bibnamefont {Schilling}},
  \bibinfo {author} {\bibfnamefont {D.~J.}\ \bibnamefont {Wilson}},\ and\
  \bibinfo {author} {\bibfnamefont {T.~J.}\ \bibnamefont {Kippenberg}},\
  }\bibfield  {title} {\bibinfo {title} {Elastic strain engineering for
  ultralow mechanical dissipation},\ }\href
  {https://www.science.org/doi/abs/10.1126/science.aar6939} {\bibfield
  {journal} {\bibinfo  {journal} {Science}\ }\textbf {\bibinfo {volume}
  {360}},\ \bibinfo {pages} {764} (\bibinfo {year} {2018})}\BibitemShut
  {NoStop}%
\bibitem [{\citenamefont {Bereyhi}\ \emph {et~al.}(2019)\citenamefont
  {Bereyhi}, \citenamefont {Beccari}, \citenamefont {Fedorov}, \citenamefont
  {Ghadimi}, \citenamefont {Schilling}, \citenamefont {Wilson}, \citenamefont
  {Engelsen},\ and\ \citenamefont {Kippenberg}}]{bereyhi_clamp-tapering_2019}%
  \BibitemOpen
  \bibfield  {author} {\bibinfo {author} {\bibfnamefont {M.~J.}\ \bibnamefont
  {Bereyhi}}, \bibinfo {author} {\bibfnamefont {A.}~\bibnamefont {Beccari}},
  \bibinfo {author} {\bibfnamefont {S.~A.}\ \bibnamefont {Fedorov}}, \bibinfo
  {author} {\bibfnamefont {A.~H.}\ \bibnamefont {Ghadimi}}, \bibinfo {author}
  {\bibfnamefont {R.}~\bibnamefont {Schilling}}, \bibinfo {author}
  {\bibfnamefont {D.~J.}\ \bibnamefont {Wilson}}, \bibinfo {author}
  {\bibfnamefont {N.~J.}\ \bibnamefont {Engelsen}},\ and\ \bibinfo {author}
  {\bibfnamefont {T.~J.}\ \bibnamefont {Kippenberg}},\ }\bibfield  {title}
  {\bibinfo {title} {Clamp-tapering increases the quality factor of stressed
  nanobeams},\ }\href {https://doi.org/10.1021/acs.nanolett.8b04942} {\bibfield
   {journal} {\bibinfo  {journal} {Nano Lett.}\ }\textbf {\bibinfo {volume}
  {19}},\ \bibinfo {pages} {2329} (\bibinfo {year} {2019})}\BibitemShut
  {NoStop}%
\bibitem [{\citenamefont {Guo}\ and\ \citenamefont
  {Gr{\"o}blacher}(2022)}]{grob_fractal_2022}%
  \BibitemOpen
  \bibfield  {author} {\bibinfo {author} {\bibfnamefont {J.}~\bibnamefont
  {Guo}}\ and\ \bibinfo {author} {\bibfnamefont {S.}~\bibnamefont
  {Gr{\"o}blacher}},\ }\bibfield  {title} {\bibinfo {title} {Integrated
  optical-readout of a high-q mechanical out-of-plane mode},\ }\href
  {https://doi.org/10.1038/s41377-022-00966-7} {\bibfield  {journal} {\bibinfo
  {journal} {Light Sci. Appl.}\ }\textbf {\bibinfo {volume} {11}},\ \bibinfo
  {pages} {282} (\bibinfo {year} {2022})}\BibitemShut {NoStop}%
\bibitem [{\citenamefont {Bereyhi}\ \emph
  {et~al.}(2022{\natexlab{a}})\citenamefont {Bereyhi}, \citenamefont {Beccari},
  \citenamefont {Groth}, \citenamefont {Fedorov}, \citenamefont {Arabmoheghi},
  \citenamefont {Kippenberg},\ and\ \citenamefont
  {Engelsen}}]{kippenberg_fractal_2021}%
  \BibitemOpen
  \bibfield  {author} {\bibinfo {author} {\bibfnamefont {M.~J.}\ \bibnamefont
  {Bereyhi}}, \bibinfo {author} {\bibfnamefont {A.}~\bibnamefont {Beccari}},
  \bibinfo {author} {\bibfnamefont {R.}~\bibnamefont {Groth}}, \bibinfo
  {author} {\bibfnamefont {S.~A.}\ \bibnamefont {Fedorov}}, \bibinfo {author}
  {\bibfnamefont {A.}~\bibnamefont {Arabmoheghi}}, \bibinfo {author}
  {\bibfnamefont {T.~J.}\ \bibnamefont {Kippenberg}},\ and\ \bibinfo {author}
  {\bibfnamefont {N.~J.}\ \bibnamefont {Engelsen}},\ }\bibfield  {title}
  {\bibinfo {title} {Hierarchical tensile structures with ultralow mechanical
  dissipation},\ }\href {https://doi.org/10.1038/s41467-022-30586-z} {\bibfield
   {journal} {\bibinfo  {journal} {Nat. Commun.}\ }\textbf {\bibinfo {volume}
  {13}},\ \bibinfo {pages} {3097} (\bibinfo {year}
  {2022}{\natexlab{a}})}\BibitemShut {NoStop}%
\bibitem [{\citenamefont {Zhou}\ \emph {et~al.}(2017)\citenamefont {Zhou},
  \citenamefont {St{\"o}hr},\ and\ \citenamefont
  {Yacoby}}]{zhou_scanning_2017}%
  \BibitemOpen
  \bibfield  {author} {\bibinfo {author} {\bibfnamefont {T.~X.}\ \bibnamefont
  {Zhou}}, \bibinfo {author} {\bibfnamefont {R.~J.}\ \bibnamefont
  {St{\"o}hr}},\ and\ \bibinfo {author} {\bibfnamefont {A.}~\bibnamefont
  {Yacoby}},\ }\bibfield  {title} {\bibinfo {title} {Scanning diamond nv center
  probes compatible with conventional afm technology},\ }\href
  {https://doi.org/10.1063/1.4995813} {\bibfield  {journal} {\bibinfo
  {journal} {Appl. Phys. Lett.}\ }\textbf {\bibinfo {volume} {111}},\ \bibinfo
  {pages} {163106} (\bibinfo {year} {2017})}\BibitemShut {NoStop}%
\bibitem [{\citenamefont {Maletinsky}\ \emph {et~al.}(2012)\citenamefont
  {Maletinsky}, \citenamefont {Hong}, \citenamefont {Grinolds}, \citenamefont
  {Hausmann}, \citenamefont {Lukin}, \citenamefont {Walsworth}, \citenamefont
  {Loncar},\ and\ \citenamefont {Yacoby}}]{maletinsky_robust_2012}%
  \BibitemOpen
  \bibfield  {author} {\bibinfo {author} {\bibfnamefont {P.}~\bibnamefont
  {Maletinsky}}, \bibinfo {author} {\bibfnamefont {S.}~\bibnamefont {Hong}},
  \bibinfo {author} {\bibfnamefont {M.~S.}\ \bibnamefont {Grinolds}}, \bibinfo
  {author} {\bibfnamefont {B.}~\bibnamefont {Hausmann}}, \bibinfo {author}
  {\bibfnamefont {M.~D.}\ \bibnamefont {Lukin}}, \bibinfo {author}
  {\bibfnamefont {R.~L.}\ \bibnamefont {Walsworth}}, \bibinfo {author}
  {\bibfnamefont {M.}~\bibnamefont {Loncar}},\ and\ \bibinfo {author}
  {\bibfnamefont {A.}~\bibnamefont {Yacoby}},\ }\bibfield  {title} {\bibinfo
  {title} {A robust scanning diamond sensor for nanoscale imaging with single
  nitrogen-vacancy centres},\ }\href {https://doi.org/10.1038/nnano.2012.50}
  {\bibfield  {journal} {\bibinfo  {journal} {Nat. Nanotechnol.}\ }\textbf
  {\bibinfo {volume} {7}},\ \bibinfo {pages} {320} (\bibinfo {year}
  {2012})}\BibitemShut {NoStop}%
\bibitem [{\citenamefont {Bluvstein}\ \emph {et~al.}(2022)\citenamefont
  {Bluvstein}, \citenamefont {Levine}, \citenamefont {Semeghini}, \citenamefont
  {Wang}, \citenamefont {Ebadi}, \citenamefont {Kalinowski}, \citenamefont
  {Keesling}, \citenamefont {Maskara}, \citenamefont {Pichler}, \citenamefont
  {Greiner}, \citenamefont {Vuleti{\'c}},\ and\ \citenamefont
  {Lukin}}]{dolev_transport}%
  \BibitemOpen
  \bibfield  {author} {\bibinfo {author} {\bibfnamefont {D.}~\bibnamefont
  {Bluvstein}}, \bibinfo {author} {\bibfnamefont {H.}~\bibnamefont {Levine}},
  \bibinfo {author} {\bibfnamefont {G.}~\bibnamefont {Semeghini}}, \bibinfo
  {author} {\bibfnamefont {T.~T.}\ \bibnamefont {Wang}}, \bibinfo {author}
  {\bibfnamefont {S.}~\bibnamefont {Ebadi}}, \bibinfo {author} {\bibfnamefont
  {M.}~\bibnamefont {Kalinowski}}, \bibinfo {author} {\bibfnamefont
  {A.}~\bibnamefont {Keesling}}, \bibinfo {author} {\bibfnamefont
  {N.}~\bibnamefont {Maskara}}, \bibinfo {author} {\bibfnamefont
  {H.}~\bibnamefont {Pichler}}, \bibinfo {author} {\bibfnamefont
  {M.}~\bibnamefont {Greiner}}, \bibinfo {author} {\bibfnamefont
  {V.}~\bibnamefont {Vuleti{\'c}}},\ and\ \bibinfo {author} {\bibfnamefont
  {M.~D.}\ \bibnamefont {Lukin}},\ }\bibfield  {title} {\bibinfo {title} {A
  quantum processor based on coherent transport of entangled atom arrays},\
  }\href {https://doi.org/10.1038/s41586-022-04592-6} {\bibfield  {journal}
  {\bibinfo  {journal} {Nature}\ }\textbf {\bibinfo {volume} {604}},\ \bibinfo
  {pages} {451–456} (\bibinfo {year} {2022})}\BibitemShut {NoStop}%
\bibitem [{\citenamefont {{\DH}or{\dj}evi{\'c}}\ \emph
  {et~al.}(2021)\citenamefont {{\DH}or{\dj}evi{\'c}}, \citenamefont
  {Samutpraphoot}, \citenamefont {Ocola}, \citenamefont {Bernien},
  \citenamefont {Grinkemeyer}, \citenamefont {Dimitrova}, \citenamefont
  {Vuleti{\'c}},\ and\ \citenamefont {Lukin}}]{dhordjevic2021entanglement}%
  \BibitemOpen
  \bibfield  {author} {\bibinfo {author} {\bibfnamefont {T.}~\bibnamefont
  {{\DH}or{\dj}evi{\'c}}}, \bibinfo {author} {\bibfnamefont {P.}~\bibnamefont
  {Samutpraphoot}}, \bibinfo {author} {\bibfnamefont {P.~L.}\ \bibnamefont
  {Ocola}}, \bibinfo {author} {\bibfnamefont {H.}~\bibnamefont {Bernien}},
  \bibinfo {author} {\bibfnamefont {B.}~\bibnamefont {Grinkemeyer}}, \bibinfo
  {author} {\bibfnamefont {I.}~\bibnamefont {Dimitrova}}, \bibinfo {author}
  {\bibfnamefont {V.}~\bibnamefont {Vuleti{\'c}}},\ and\ \bibinfo {author}
  {\bibfnamefont {M.~D.}\ \bibnamefont {Lukin}},\ }\bibfield  {title} {\bibinfo
  {title} {Entanglement transport and a nanophotonic interface for atoms in
  optical tweezers},\ }\href {https://doi.org/10.1126/science.abi9917}
  {\bibfield  {journal} {\bibinfo  {journal} {Science}\ }\textbf {\bibinfo
  {volume} {373}},\ \bibinfo {pages} {1511} (\bibinfo {year}
  {2021})}\BibitemShut {NoStop}%
\bibitem [{\citenamefont {Mandel}\ \emph {et~al.}(2003)\citenamefont {Mandel},
  \citenamefont {Greiner}, \citenamefont {Widera}, \citenamefont {Rom},
  \citenamefont {H{\"a}nsch},\ and\ \citenamefont
  {Bloch}}]{mandel2003coherent}%
  \BibitemOpen
  \bibfield  {author} {\bibinfo {author} {\bibfnamefont {O.}~\bibnamefont
  {Mandel}}, \bibinfo {author} {\bibfnamefont {M.}~\bibnamefont {Greiner}},
  \bibinfo {author} {\bibfnamefont {A.}~\bibnamefont {Widera}}, \bibinfo
  {author} {\bibfnamefont {T.}~\bibnamefont {Rom}}, \bibinfo {author}
  {\bibfnamefont {T.~W.}\ \bibnamefont {H{\"a}nsch}},\ and\ \bibinfo {author}
  {\bibfnamefont {I.}~\bibnamefont {Bloch}},\ }\bibfield  {title} {\bibinfo
  {title} {Coherent transport of neutral atoms in spin-dependent optical
  lattice potentials},\ }\href
  {https://link.aps.org/doi/10.1103/PhysRevLett.91.010407} {\bibfield
  {journal} {\bibinfo  {journal} {Phys. Rev. Lett.}\ }\textbf {\bibinfo
  {volume} {91}},\ \bibinfo {pages} {010407} (\bibinfo {year}
  {2003})}\BibitemShut {NoStop}%
\bibitem [{\citenamefont {Pino}\ \emph {et~al.}(2021)\citenamefont {Pino},
  \citenamefont {Dreiling}, \citenamefont {Figgatt}, \citenamefont {Gaebler},
  \citenamefont {Moses}, \citenamefont {Allman}, \citenamefont {Baldwin},
  \citenamefont {Foss-Feig}, \citenamefont {Hayes}, \citenamefont {Mayer} \emph
  {et~al.}}]{pino2021demonstration}%
  \BibitemOpen
  \bibfield  {author} {\bibinfo {author} {\bibfnamefont {J.~M.}\ \bibnamefont
  {Pino}}, \bibinfo {author} {\bibfnamefont {J.~M.}\ \bibnamefont {Dreiling}},
  \bibinfo {author} {\bibfnamefont {C.}~\bibnamefont {Figgatt}}, \bibinfo
  {author} {\bibfnamefont {J.~P.}\ \bibnamefont {Gaebler}}, \bibinfo {author}
  {\bibfnamefont {S.~A.}\ \bibnamefont {Moses}}, \bibinfo {author}
  {\bibfnamefont {M.~S.}\ \bibnamefont {Allman}}, \bibinfo {author}
  {\bibfnamefont {C.~H.}\ \bibnamefont {Baldwin}}, \bibinfo {author}
  {\bibfnamefont {M.}~\bibnamefont {Foss-Feig}}, \bibinfo {author}
  {\bibfnamefont {D.}~\bibnamefont {Hayes}}, \bibinfo {author} {\bibfnamefont
  {K.}~\bibnamefont {Mayer}}, \emph {et~al.},\ }\bibfield  {title} {\bibinfo
  {title} {Demonstration of the trapped-ion quantum ccd computer
  architecture},\ }\href {https://doi.org/10.1038/s41586-021-03318-4}
  {\bibfield  {journal} {\bibinfo  {journal} {Nature}\ }\textbf {\bibinfo
  {volume} {592}},\ \bibinfo {pages} {209} (\bibinfo {year}
  {2021})}\BibitemShut {NoStop}%
\bibitem [{\citenamefont {Monroe}\ \emph {et~al.}(2014)\citenamefont {Monroe},
  \citenamefont {Raussendorf}, \citenamefont {Ruthven}, \citenamefont {Brown},
  \citenamefont {Maunz}, \citenamefont {Duan},\ and\ \citenamefont
  {Kim}}]{monroe2014large}%
  \BibitemOpen
  \bibfield  {author} {\bibinfo {author} {\bibfnamefont {C.}~\bibnamefont
  {Monroe}}, \bibinfo {author} {\bibfnamefont {R.}~\bibnamefont {Raussendorf}},
  \bibinfo {author} {\bibfnamefont {A.}~\bibnamefont {Ruthven}}, \bibinfo
  {author} {\bibfnamefont {K.~R.}\ \bibnamefont {Brown}}, \bibinfo {author}
  {\bibfnamefont {P.}~\bibnamefont {Maunz}}, \bibinfo {author} {\bibfnamefont
  {L.-M.}\ \bibnamefont {Duan}},\ and\ \bibinfo {author} {\bibfnamefont
  {J.}~\bibnamefont {Kim}},\ }\bibfield  {title} {\bibinfo {title} {Large-scale
  modular quantum-computer architecture with atomic memory and photonic
  interconnects},\ }\href {https://link.aps.org/doi/10.1103/PhysRevA.89.022317}
  {\bibfield  {journal} {\bibinfo  {journal} {Phys. Rev. A}\ }\textbf {\bibinfo
  {volume} {89}},\ \bibinfo {pages} {022317} (\bibinfo {year}
  {2014})}\BibitemShut {NoStop}%
\bibitem [{\citenamefont {Cirac}\ and\ \citenamefont
  {Zoller}(2000)}]{cirac2000scalable}%
  \BibitemOpen
  \bibfield  {author} {\bibinfo {author} {\bibfnamefont {J.~I.}\ \bibnamefont
  {Cirac}}\ and\ \bibinfo {author} {\bibfnamefont {P.}~\bibnamefont {Zoller}},\
  }\bibfield  {title} {\bibinfo {title} {A scalable quantum computer with ions
  in an array of microtraps},\ }\href {https://doi.org/10.1038/35007021}
  {\bibfield  {journal} {\bibinfo  {journal} {Nature}\ }\textbf {\bibinfo
  {volume} {404}},\ \bibinfo {pages} {579} (\bibinfo {year}
  {2000})}\BibitemShut {NoStop}%
\bibitem [{\citenamefont {Xie}\ \emph {et~al.}(2018)\citenamefont {Xie},
  \citenamefont {Zhou}, \citenamefont {St\"{o}hr},\ and\ \citenamefont
  {Yacoby}}]{angle_etch}%
  \BibitemOpen
  \bibfield  {author} {\bibinfo {author} {\bibfnamefont {L.}~\bibnamefont
  {Xie}}, \bibinfo {author} {\bibfnamefont {T.~X.}\ \bibnamefont {Zhou}},
  \bibinfo {author} {\bibfnamefont {R.~J.}\ \bibnamefont {St\"{o}hr}},\ and\
  \bibinfo {author} {\bibfnamefont {A.}~\bibnamefont {Yacoby}},\ }\bibfield
  {title} {\bibinfo {title} {Crystallographic orientation dependent reactive
  ion etching in single crystal diamond},\ }\href
  {https://doi.org/https://doi-org.ezp-prod1.hul.harvard.edu/10.1002/adma.201705501}
  {\bibfield  {journal} {\bibinfo  {journal} {Adv. Mat.}\ }\textbf {\bibinfo
  {volume} {30}},\ \bibinfo {pages} {1705501} (\bibinfo {year}
  {2018})}\BibitemShut {NoStop}%
\bibitem [{sup()}]{supp}%
  \BibitemOpen
  \href@noop {} {}\bibinfo {note} {See Supplemental Material.}\BibitemShut
  {Stop}%
\bibitem [{\citenamefont {Stanwix}\ \emph {et~al.}(2010)\citenamefont
  {Stanwix}, \citenamefont {Pham}, \citenamefont {Maze}, \citenamefont
  {Le~Sage}, \citenamefont {Yeung}, \citenamefont {Cappellaro}, \citenamefont
  {Hemmer}, \citenamefont {Yacoby}, \citenamefont {Lukin},\ and\ \citenamefont
  {Walsworth}}]{stanwix_coherence_2010}%
  \BibitemOpen
  \bibfield  {author} {\bibinfo {author} {\bibfnamefont {P.~L.}\ \bibnamefont
  {Stanwix}}, \bibinfo {author} {\bibfnamefont {L.~M.}\ \bibnamefont {Pham}},
  \bibinfo {author} {\bibfnamefont {J.~R.}\ \bibnamefont {Maze}}, \bibinfo
  {author} {\bibfnamefont {D.}~\bibnamefont {Le~Sage}}, \bibinfo {author}
  {\bibfnamefont {T.~K.}\ \bibnamefont {Yeung}}, \bibinfo {author}
  {\bibfnamefont {P.}~\bibnamefont {Cappellaro}}, \bibinfo {author}
  {\bibfnamefont {P.~R.}\ \bibnamefont {Hemmer}}, \bibinfo {author}
  {\bibfnamefont {A.}~\bibnamefont {Yacoby}}, \bibinfo {author} {\bibfnamefont
  {M.~D.}\ \bibnamefont {Lukin}},\ and\ \bibinfo {author} {\bibfnamefont
  {R.~L.}\ \bibnamefont {Walsworth}},\ }\bibfield  {title} {\bibinfo {title}
  {Coherence of nitrogen-vacancy electronic spin ensembles in diamond},\ }\href
  {https://link.aps.org/doi/10.1103/PhysRevB.82.201201} {\bibfield  {journal}
  {\bibinfo  {journal} {Phys. Rev. B}\ }\textbf {\bibinfo {volume} {82}},\
  \bibinfo {pages} {201201} (\bibinfo {year} {2010})}\BibitemShut {NoStop}%
\bibitem [{\citenamefont {Pfender}\ \emph
  {et~al.}(2017{\natexlab{a}})\citenamefont {Pfender}, \citenamefont {Aslam},
  \citenamefont {Sumiya}, \citenamefont {Onoda}, \citenamefont {Neumann},
  \citenamefont {Isoya}, \citenamefont {Meriles},\ and\ \citenamefont
  {Wrachtrup}}]{pfender2017nonvolatile}%
  \BibitemOpen
  \bibfield  {author} {\bibinfo {author} {\bibfnamefont {M.}~\bibnamefont
  {Pfender}}, \bibinfo {author} {\bibfnamefont {N.}~\bibnamefont {Aslam}},
  \bibinfo {author} {\bibfnamefont {H.}~\bibnamefont {Sumiya}}, \bibinfo
  {author} {\bibfnamefont {S.}~\bibnamefont {Onoda}}, \bibinfo {author}
  {\bibfnamefont {P.}~\bibnamefont {Neumann}}, \bibinfo {author} {\bibfnamefont
  {J.}~\bibnamefont {Isoya}}, \bibinfo {author} {\bibfnamefont {C.~A.}\
  \bibnamefont {Meriles}},\ and\ \bibinfo {author} {\bibfnamefont
  {J.}~\bibnamefont {Wrachtrup}},\ }\bibfield  {title} {\bibinfo {title}
  {Nonvolatile nuclear spin memory enables sensor-unlimited nanoscale
  spectroscopy of small spin clusters},\ }\href
  {https://doi.org/10.1038/s41467-017-00964-z} {\bibfield  {journal} {\bibinfo
  {journal} {Nat. Commun.}\ }\textbf {\bibinfo {volume} {8}},\ \bibinfo {pages}
  {834} (\bibinfo {year} {2017}{\natexlab{a}})}\BibitemShut {NoStop}%
\bibitem [{\citenamefont {Zaiser}\ \emph {et~al.}(2016)\citenamefont {Zaiser},
  \citenamefont {Rendler}, \citenamefont {Jakobi}, \citenamefont {Wolf},
  \citenamefont {Lee}, \citenamefont {Wagner}, \citenamefont {Bergholm},
  \citenamefont {Schulte-Herbr{\"u}ggen}, \citenamefont {Neumann},\ and\
  \citenamefont {Wrachtrup}}]{zaiser2016enhancing}%
  \BibitemOpen
  \bibfield  {author} {\bibinfo {author} {\bibfnamefont {S.}~\bibnamefont
  {Zaiser}}, \bibinfo {author} {\bibfnamefont {T.}~\bibnamefont {Rendler}},
  \bibinfo {author} {\bibfnamefont {I.}~\bibnamefont {Jakobi}}, \bibinfo
  {author} {\bibfnamefont {T.}~\bibnamefont {Wolf}}, \bibinfo {author}
  {\bibfnamefont {S.-Y.}\ \bibnamefont {Lee}}, \bibinfo {author} {\bibfnamefont
  {S.}~\bibnamefont {Wagner}}, \bibinfo {author} {\bibfnamefont
  {V.}~\bibnamefont {Bergholm}}, \bibinfo {author} {\bibfnamefont
  {T.}~\bibnamefont {Schulte-Herbr{\"u}ggen}}, \bibinfo {author} {\bibfnamefont
  {P.}~\bibnamefont {Neumann}},\ and\ \bibinfo {author} {\bibfnamefont
  {J.}~\bibnamefont {Wrachtrup}},\ }\bibfield  {title} {\bibinfo {title}
  {Enhancing quantum sensing sensitivity by a quantum memory},\ }\href
  {https://doi.org/10.1038/ncomms12279} {\bibfield  {journal} {\bibinfo
  {journal} {Nat. Commun.}\ }\textbf {\bibinfo {volume} {7}},\ \bibinfo {pages}
  {12279} (\bibinfo {year} {2016})}\BibitemShut {NoStop}%
\bibitem [{\citenamefont {Jiang}\ \emph {et~al.}(2009)\citenamefont {Jiang},
  \citenamefont {Hodges}, \citenamefont {Maze}, \citenamefont {Maurer},
  \citenamefont {Taylor}, \citenamefont {Cory}, \citenamefont {Hemmer},
  \citenamefont {Walsworth}, \citenamefont {Yacoby}, \citenamefont {Zibrov},\
  and\ \citenamefont {Lukin}}]{jiang2009repetitive}%
  \BibitemOpen
  \bibfield  {author} {\bibinfo {author} {\bibfnamefont {L.}~\bibnamefont
  {Jiang}}, \bibinfo {author} {\bibfnamefont {J.~S.}\ \bibnamefont {Hodges}},
  \bibinfo {author} {\bibfnamefont {J.~R.}\ \bibnamefont {Maze}}, \bibinfo
  {author} {\bibfnamefont {P.}~\bibnamefont {Maurer}}, \bibinfo {author}
  {\bibfnamefont {J.~M.}\ \bibnamefont {Taylor}}, \bibinfo {author}
  {\bibfnamefont {D.~G.}\ \bibnamefont {Cory}}, \bibinfo {author}
  {\bibfnamefont {P.~R.}\ \bibnamefont {Hemmer}}, \bibinfo {author}
  {\bibfnamefont {R.~L.}\ \bibnamefont {Walsworth}}, \bibinfo {author}
  {\bibfnamefont {A.}~\bibnamefont {Yacoby}}, \bibinfo {author} {\bibfnamefont
  {A.~S.}\ \bibnamefont {Zibrov}},\ and\ \bibinfo {author} {\bibfnamefont
  {M.~D.}\ \bibnamefont {Lukin}},\ }\bibfield  {title} {\bibinfo {title}
  {Repetitive readout of a single electronic spin via quantum logic with
  nuclear spin ancillae},\ }\href {https://doi.org/10.1126/science.1176496}
  {\bibfield  {journal} {\bibinfo  {journal} {Science}\ }\textbf {\bibinfo
  {volume} {326}},\ \bibinfo {pages} {267} (\bibinfo {year}
  {2009})}\BibitemShut {NoStop}%
\bibitem [{\citenamefont {Pfender}\ \emph
  {et~al.}(2017{\natexlab{b}})\citenamefont {Pfender}, \citenamefont {Aslam},
  \citenamefont {Simon}, \citenamefont {Antonov}, \citenamefont {Thiering},
  \citenamefont {Burk}, \citenamefont {F{\'a}varo~de Oliveira}, \citenamefont
  {Denisenko}, \citenamefont {Fedder}, \citenamefont {Meijer},\ and\
  \citenamefont {et~al.}}]{pfender2017protecting}%
  \BibitemOpen
  \bibfield  {author} {\bibinfo {author} {\bibfnamefont {M.}~\bibnamefont
  {Pfender}}, \bibinfo {author} {\bibfnamefont {N.}~\bibnamefont {Aslam}},
  \bibinfo {author} {\bibfnamefont {P.}~\bibnamefont {Simon}}, \bibinfo
  {author} {\bibfnamefont {D.}~\bibnamefont {Antonov}}, \bibinfo {author}
  {\bibfnamefont {G.}~\bibnamefont {Thiering}}, \bibinfo {author}
  {\bibfnamefont {S.}~\bibnamefont {Burk}}, \bibinfo {author} {\bibfnamefont
  {F.}~\bibnamefont {F{\'a}varo~de Oliveira}}, \bibinfo {author} {\bibfnamefont
  {A.}~\bibnamefont {Denisenko}}, \bibinfo {author} {\bibfnamefont
  {H.}~\bibnamefont {Fedder}}, \bibinfo {author} {\bibfnamefont
  {J.}~\bibnamefont {Meijer}},\ and\ \bibinfo {author} {\bibnamefont
  {et~al.}},\ }\bibfield  {title} {\bibinfo {title} {Protecting a diamond
  quantum memory by charge state control},\ }\href
  {https://doi.org/10.1021/acs.nanolett.7b01796} {\bibfield  {journal}
  {\bibinfo  {journal} {Nano Lett.}\ }\textbf {\bibinfo {volume} {17}},\
  \bibinfo {pages} {5931} (\bibinfo {year} {2017}{\natexlab{b}})}\BibitemShut
  {NoStop}%
\bibitem [{\citenamefont {Vinante}\ \emph {et~al.}(2016)\citenamefont
  {Vinante}, \citenamefont {Bahrami}, \citenamefont {Bassi}, \citenamefont
  {Usenko}, \citenamefont {Wijts},\ and\ \citenamefont
  {Oosterkamp}}]{vinante_upper_2016}%
  \BibitemOpen
  \bibfield  {author} {\bibinfo {author} {\bibfnamefont {A.}~\bibnamefont
  {Vinante}}, \bibinfo {author} {\bibfnamefont {M.}~\bibnamefont {Bahrami}},
  \bibinfo {author} {\bibfnamefont {A.}~\bibnamefont {Bassi}}, \bibinfo
  {author} {\bibfnamefont {O.}~\bibnamefont {Usenko}}, \bibinfo {author}
  {\bibfnamefont {G.}~\bibnamefont {Wijts}},\ and\ \bibinfo {author}
  {\bibfnamefont {T.~H.}\ \bibnamefont {Oosterkamp}},\ }\bibfield  {title}
  {\bibinfo {title} {Upper bounds on spontaneous wave-function collapse models
  using millikelvin-cooled nanocantilevers},\ }\href
  {https://doi.org/10.1103/PhysRevLett.116.090402} {\bibfield  {journal}
  {\bibinfo  {journal} {Phys. Rev. Lett.}\ }\textbf {\bibinfo {volume} {116}},\
  \bibinfo {pages} {090402} (\bibinfo {year} {2016})}\BibitemShut {NoStop}%
\bibitem [{\citenamefont {Van~Wezel}\ and\ \citenamefont
  {Oosterkamp}(2012)}]{van_wezel_nanoscale_2012}%
  \BibitemOpen
  \bibfield  {author} {\bibinfo {author} {\bibfnamefont {J.}~\bibnamefont
  {Van~Wezel}}\ and\ \bibinfo {author} {\bibfnamefont {T.~H.}\ \bibnamefont
  {Oosterkamp}},\ }\bibfield  {title} {\bibinfo {title} {A nanoscale experiment
  measuring gravity's role in breaking the unitarity of quantum dynamics},\
  }\href {https://doi.org/10.1098/rspa.2011.0201} {\bibfield  {journal}
  {\bibinfo  {journal} {Proc. R. Soc. A.}\ }\textbf {\bibinfo {volume} {468}},\
  \bibinfo {pages} {35} (\bibinfo {year} {2012})}\BibitemShut {NoStop}%
\bibitem [{\citenamefont {Ariyaratne}\ \emph {et~al.}(2018)\citenamefont
  {Ariyaratne}, \citenamefont {Bluvstein}, \citenamefont {Myers},\ and\
  \citenamefont {Jayich}}]{jayich_afm}%
  \BibitemOpen
  \bibfield  {author} {\bibinfo {author} {\bibfnamefont {A.}~\bibnamefont
  {Ariyaratne}}, \bibinfo {author} {\bibfnamefont {D.}~\bibnamefont
  {Bluvstein}}, \bibinfo {author} {\bibfnamefont {B.~A.}\ \bibnamefont
  {Myers}},\ and\ \bibinfo {author} {\bibfnamefont {A.~C.~B.}\ \bibnamefont
  {Jayich}},\ }\bibfield  {title} {\bibinfo {title} {Nanoscale electrical
  conductivity imaging using a nitrogen-vacancy center in diamond},\ }\href
  {https://doi.org/10.1038/s41467-018-04798-1} {\bibfield  {journal} {\bibinfo
  {journal} {Nat. Commun.}\ }\textbf {\bibinfo {volume} {9}},\ \bibinfo {pages}
  {2406} (\bibinfo {year} {2018})}\BibitemShut {NoStop}%
\bibitem [{\citenamefont {Bereyhi}\ \emph
  {et~al.}(2022{\natexlab{b}})\citenamefont {Bereyhi}, \citenamefont
  {Arabmoheghi}, \citenamefont {Beccari}, \citenamefont {Fedorov},
  \citenamefont {Huang}, \citenamefont {Kippenberg},\ and\ \citenamefont
  {Engelsen}}]{bereyhi2022perimeter}%
  \BibitemOpen
  \bibfield  {author} {\bibinfo {author} {\bibfnamefont {M.~J.}\ \bibnamefont
  {Bereyhi}}, \bibinfo {author} {\bibfnamefont {A.}~\bibnamefont
  {Arabmoheghi}}, \bibinfo {author} {\bibfnamefont {A.}~\bibnamefont
  {Beccari}}, \bibinfo {author} {\bibfnamefont {S.~A.}\ \bibnamefont
  {Fedorov}}, \bibinfo {author} {\bibfnamefont {G.}~\bibnamefont {Huang}},
  \bibinfo {author} {\bibfnamefont {T.~J.}\ \bibnamefont {Kippenberg}},\ and\
  \bibinfo {author} {\bibfnamefont {N.~J.}\ \bibnamefont {Engelsen}},\
  }\bibfield  {title} {\bibinfo {title} {Perimeter modes of nanomechanical
  resonators exhibit quality factors exceeding 10 9 at room temperature},\
  }\href {https://link.aps.org/doi/10.1103/PhysRevX.12.021036} {\bibfield
  {journal} {\bibinfo  {journal} {Phys. Rev. X}\ }\textbf {\bibinfo {volume}
  {12}},\ \bibinfo {pages} {021036} (\bibinfo {year}
  {2022}{\natexlab{b}})}\BibitemShut {NoStop}%
\bibitem [{\citenamefont {Beccari}\ \emph {et~al.}(2022)\citenamefont
  {Beccari}, \citenamefont {Visani}, \citenamefont {Fedorov}, \citenamefont
  {Bereyhi}, \citenamefont {Boureau}, \citenamefont {Engelsen},\ and\
  \citenamefont {Kippenberg}}]{beccari2022strained}%
  \BibitemOpen
  \bibfield  {author} {\bibinfo {author} {\bibfnamefont {A.}~\bibnamefont
  {Beccari}}, \bibinfo {author} {\bibfnamefont {D.~A.}\ \bibnamefont {Visani}},
  \bibinfo {author} {\bibfnamefont {S.~A.}\ \bibnamefont {Fedorov}}, \bibinfo
  {author} {\bibfnamefont {M.~J.}\ \bibnamefont {Bereyhi}}, \bibinfo {author}
  {\bibfnamefont {V.}~\bibnamefont {Boureau}}, \bibinfo {author} {\bibfnamefont
  {N.~J.}\ \bibnamefont {Engelsen}},\ and\ \bibinfo {author} {\bibfnamefont
  {T.~J.}\ \bibnamefont {Kippenberg}},\ }\bibfield  {title} {\bibinfo {title}
  {Strained crystalline nanomechanical resonators with quality factors above 10
  billion},\ }\href {https://doi.org/10.1038/s41567-021-01498-4} {\bibfield
  {journal} {\bibinfo  {journal} {Nat. Phys.}\ }\textbf {\bibinfo {volume}
  {18}},\ \bibinfo {pages} {436} (\bibinfo {year} {2022})}\BibitemShut
  {NoStop}%
\bibitem [{\citenamefont {Sementilli}\ \emph {et~al.}(2022)\citenamefont
  {Sementilli}, \citenamefont {Romero},\ and\ \citenamefont
  {Bowen}}]{sementilli2022nanomechanical}%
  \BibitemOpen
  \bibfield  {author} {\bibinfo {author} {\bibfnamefont {L.}~\bibnamefont
  {Sementilli}}, \bibinfo {author} {\bibfnamefont {E.}~\bibnamefont {Romero}},\
  and\ \bibinfo {author} {\bibfnamefont {W.~P.}\ \bibnamefont {Bowen}},\
  }\bibfield  {title} {\bibinfo {title} {Nanomechanical dissipation and strain
  engineering},\ }\href {https://doi.org/10.1002/adfm.202105247} {\bibfield
  {journal} {\bibinfo  {journal} {Adv. Funct. Mater.}\ }\textbf {\bibinfo
  {volume} {32}},\ \bibinfo {pages} {2105247} (\bibinfo {year}
  {2022})}\BibitemShut {NoStop}%
\bibitem [{\citenamefont {Bar-Gill}\ \emph {et~al.}(2013)\citenamefont
  {Bar-Gill}, \citenamefont {Pham}, \citenamefont {Jarmola}, \citenamefont
  {Budker},\ and\ \citenamefont {Walsworth}}]{bar-gill_solid-state_2013}%
  \BibitemOpen
  \bibfield  {author} {\bibinfo {author} {\bibfnamefont {N.}~\bibnamefont
  {Bar-Gill}}, \bibinfo {author} {\bibfnamefont {L.~M.}\ \bibnamefont {Pham}},
  \bibinfo {author} {\bibfnamefont {A.}~\bibnamefont {Jarmola}}, \bibinfo
  {author} {\bibfnamefont {D.}~\bibnamefont {Budker}},\ and\ \bibinfo {author}
  {\bibfnamefont {R.~L.}\ \bibnamefont {Walsworth}},\ }\bibfield  {title}
  {\bibinfo {title} {Solid-state electronic spin coherence time approaching one
  second},\ }\href {https://doi.org/10.1038/ncomms2771} {\bibfield  {journal}
  {\bibinfo  {journal} {Nat. Commun.}\ }\textbf {\bibinfo {volume} {4}},\
  \bibinfo {pages} {1743} (\bibinfo {year} {2013})}\BibitemShut {NoStop}%
\bibitem [{\citenamefont {Luan}\ \emph {et~al.}(2015)\citenamefont {Luan},
  \citenamefont {Grinolds}, \citenamefont {Hong}, \citenamefont {Maletinsky},
  \citenamefont {Walsworth},\ and\ \citenamefont
  {Yacoby}}]{luan_decoherence_nodate}%
  \BibitemOpen
  \bibfield  {author} {\bibinfo {author} {\bibfnamefont {L.}~\bibnamefont
  {Luan}}, \bibinfo {author} {\bibfnamefont {M.~S.}\ \bibnamefont {Grinolds}},
  \bibinfo {author} {\bibfnamefont {S.}~\bibnamefont {Hong}}, \bibinfo {author}
  {\bibfnamefont {P.}~\bibnamefont {Maletinsky}}, \bibinfo {author}
  {\bibfnamefont {R.~L.}\ \bibnamefont {Walsworth}},\ and\ \bibinfo {author}
  {\bibfnamefont {A.}~\bibnamefont {Yacoby}},\ }\bibfield  {title} {\bibinfo
  {title} {Decoherence imaging of spin ensembles using a scanning
  single-electron spin in diamond},\ }\href {https://doi.org/10.1038/srep08119}
  {\bibfield  {journal} {\bibinfo  {journal} {Sci. Rep.}\ }\textbf {\bibinfo
  {volume} {5}},\ \bibinfo {pages} {8119} (\bibinfo {year} {2015})}\BibitemShut
  {NoStop}%
\bibitem [{\citenamefont {Sangtawesin}\ \emph {et~al.}(2019)\citenamefont
  {Sangtawesin}, \citenamefont {Dwyer}, \citenamefont {Srinivasan},
  \citenamefont {Allred}, \citenamefont {Rodgers}, \citenamefont {De~Greve},
  \citenamefont {Stacey}, \citenamefont {Dontschuk}, \citenamefont
  {O’Donnell}, \citenamefont {Hu}, \citenamefont {Evans}, \citenamefont
  {Jaye}, \citenamefont {Fischer}, \citenamefont {Markham}, \citenamefont
  {Twitchen}, \citenamefont {Park}, \citenamefont {Lukin},\ and\ \citenamefont
  {de~Leon}}]{sangtawesin_origins_2019}%
  \BibitemOpen
  \bibfield  {author} {\bibinfo {author} {\bibfnamefont {S.}~\bibnamefont
  {Sangtawesin}}, \bibinfo {author} {\bibfnamefont {B.~L.}\ \bibnamefont
  {Dwyer}}, \bibinfo {author} {\bibfnamefont {S.}~\bibnamefont {Srinivasan}},
  \bibinfo {author} {\bibfnamefont {J.~J.}\ \bibnamefont {Allred}}, \bibinfo
  {author} {\bibfnamefont {L.~V.~H.}\ \bibnamefont {Rodgers}}, \bibinfo
  {author} {\bibfnamefont {K.}~\bibnamefont {De~Greve}}, \bibinfo {author}
  {\bibfnamefont {A.}~\bibnamefont {Stacey}}, \bibinfo {author} {\bibfnamefont
  {N.}~\bibnamefont {Dontschuk}}, \bibinfo {author} {\bibfnamefont {K.~M.}\
  \bibnamefont {O’Donnell}}, \bibinfo {author} {\bibfnamefont
  {D.}~\bibnamefont {Hu}}, \bibinfo {author} {\bibfnamefont {D.~A.}\
  \bibnamefont {Evans}}, \bibinfo {author} {\bibfnamefont {C.}~\bibnamefont
  {Jaye}}, \bibinfo {author} {\bibfnamefont {D.~A.}\ \bibnamefont {Fischer}},
  \bibinfo {author} {\bibfnamefont {M.~L.}\ \bibnamefont {Markham}}, \bibinfo
  {author} {\bibfnamefont {D.~J.}\ \bibnamefont {Twitchen}}, \bibinfo {author}
  {\bibfnamefont {H.}~\bibnamefont {Park}}, \bibinfo {author} {\bibfnamefont
  {M.~D.}\ \bibnamefont {Lukin}},\ and\ \bibinfo {author} {\bibfnamefont
  {N.~P.}\ \bibnamefont {de~Leon}},\ }\bibfield  {title} {\bibinfo {title}
  {Origins of diamond surface noise probed by correlating single-spin
  measurements with surface spectroscopy},\ }\href
  {https://doi.org/10.1103/PhysRevX.9.031052} {\bibfield  {journal} {\bibinfo
  {journal} {Phys. Rev. X}\ }\textbf {\bibinfo {volume} {9}},\ \bibinfo {pages}
  {031052} (\bibinfo {year} {2019})}\BibitemShut {NoStop}%
\bibitem [{\citenamefont {Herbschleb}\ \emph {et~al.}(2019)\citenamefont
  {Herbschleb}, \citenamefont {Kato}, \citenamefont {Maruyama}, \citenamefont
  {Danjo}, \citenamefont {Makino}, \citenamefont {Yamasaki}, \citenamefont
  {Ohki}, \citenamefont {Hayashi}, \citenamefont {Morishita}, \citenamefont
  {Fujiwara} \emph {et~al.}}]{herbschleb2019ultra}%
  \BibitemOpen
  \bibfield  {author} {\bibinfo {author} {\bibfnamefont {E.~D.}\ \bibnamefont
  {Herbschleb}}, \bibinfo {author} {\bibfnamefont {H.}~\bibnamefont {Kato}},
  \bibinfo {author} {\bibfnamefont {Y.}~\bibnamefont {Maruyama}}, \bibinfo
  {author} {\bibfnamefont {T.}~\bibnamefont {Danjo}}, \bibinfo {author}
  {\bibfnamefont {T.}~\bibnamefont {Makino}}, \bibinfo {author} {\bibfnamefont
  {S.}~\bibnamefont {Yamasaki}}, \bibinfo {author} {\bibfnamefont
  {I.}~\bibnamefont {Ohki}}, \bibinfo {author} {\bibfnamefont {K.}~\bibnamefont
  {Hayashi}}, \bibinfo {author} {\bibfnamefont {H.}~\bibnamefont {Morishita}},
  \bibinfo {author} {\bibfnamefont {M.}~\bibnamefont {Fujiwara}}, \emph
  {et~al.},\ }\bibfield  {title} {\bibinfo {title} {Ultra-long coherence times
  amongst room-temperature solid-state spins},\ }\href
  {https://doi.org/10.1038/s41467-019-11776-8} {\bibfield  {journal} {\bibinfo
  {journal} {Nat. Commun.}\ }\textbf {\bibinfo {volume} {10}},\ \bibinfo
  {pages} {3766} (\bibinfo {year} {2019})}\BibitemShut {NoStop}%
\bibitem [{\citenamefont {Clerk}\ \emph {et~al.}(2020)\citenamefont {Clerk},
  \citenamefont {Lehnert}, \citenamefont {Bertet}, \citenamefont {Petta},\ and\
  \citenamefont {Nakamura}}]{clerk2020hybrid}%
  \BibitemOpen
  \bibfield  {author} {\bibinfo {author} {\bibfnamefont {A.~A.}\ \bibnamefont
  {Clerk}}, \bibinfo {author} {\bibfnamefont {K.~W.}\ \bibnamefont {Lehnert}},
  \bibinfo {author} {\bibfnamefont {P.}~\bibnamefont {Bertet}}, \bibinfo
  {author} {\bibfnamefont {J.~R.}\ \bibnamefont {Petta}},\ and\ \bibinfo
  {author} {\bibfnamefont {Y.}~\bibnamefont {Nakamura}},\ }\bibfield  {title}
  {\bibinfo {title} {Hybrid quantum systems with circuit quantum
  electrodynamics},\ }\href {https://doi.org/10.1038/s41567-020-0797-9}
  {\bibfield  {journal} {\bibinfo  {journal} {Nat. Phys.}\ }\textbf {\bibinfo
  {volume} {16}},\ \bibinfo {pages} {257} (\bibinfo {year} {2020})}\BibitemShut
  {NoStop}%
\bibitem [{\citenamefont {Barzanjeh}\ \emph {et~al.}(2022)\citenamefont
  {Barzanjeh}, \citenamefont {Xuereb}, \citenamefont {Gr{\"o}blacher},
  \citenamefont {Paternostro}, \citenamefont {Regal},\ and\ \citenamefont
  {Weig}}]{barzanjeh2022optomechanics}%
  \BibitemOpen
  \bibfield  {author} {\bibinfo {author} {\bibfnamefont {S.}~\bibnamefont
  {Barzanjeh}}, \bibinfo {author} {\bibfnamefont {A.}~\bibnamefont {Xuereb}},
  \bibinfo {author} {\bibfnamefont {S.}~\bibnamefont {Gr{\"o}blacher}},
  \bibinfo {author} {\bibfnamefont {M.}~\bibnamefont {Paternostro}}, \bibinfo
  {author} {\bibfnamefont {C.~A.}\ \bibnamefont {Regal}},\ and\ \bibinfo
  {author} {\bibfnamefont {E.~M.}\ \bibnamefont {Weig}},\ }\bibfield  {title}
  {\bibinfo {title} {Optomechanics for quantum technologies},\ }\href
  {https://doi.org/10.1038/s41567-021-01402-0} {\bibfield  {journal} {\bibinfo
  {journal} {Nat. Phys.}\ }\textbf {\bibinfo {volume} {18}},\ \bibinfo {pages}
  {15} (\bibinfo {year} {2022})}\BibitemShut {NoStop}%
\end{thebibliography}%

%\onecolumngrid
%\clearpage
%\include{SI_strings.tex}

\end{document}

\include{commands}

\title{\Supplementary Information for \\
Programmable Quantum Processors based on Spin Qubits with Mechanically-Mediated Interactions and Transport}
\author{}
\maketitle

\tableofcontents

\section{Experimental setup}
The experiment is performed inside a modified Janis ST-500 continuous flow cryostat, which also serves as a vacuum chamber for room temperature measurements shown in Fig.~2 and 3. The AFM chip containing the diamond nanopillar is attached to a glass slide which is then mounted on a fixed copper block. The sample containing the nanobeams is mounted on a 3-axis Attocube stack (ANPz101 and ANPx101), which allows for fine control over the position of the diamond nanopillar relative to nanobeam.

\begin{figure}[!htb]
\centering
\includegraphics[width=.6\linewidth]{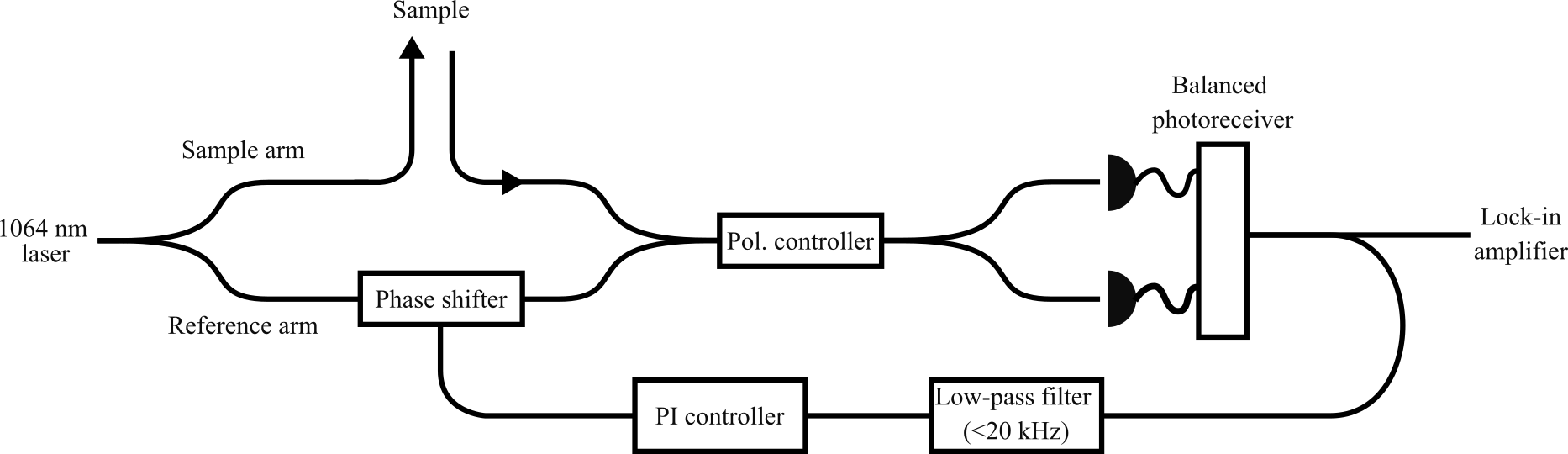}
\caption{Schematic of the interferometer. A 1064 nm laser is split into a sample and reference arm. Power in the sample arm is reflected off a nanobeam and then recombined with the reference arm. The light is then split off again by a polarization beam splitter to be detected by a balanced photoreceiver (Newport 2117-FC), where the output is sent for lock-in detection (Zurich Instruments HF2) and also used to phase-stabilize the reference arm against low-frequency noise.}
\label{fig:interferometer}
\end{figure}

We use a home-built confocal microscope to perform NV measurements. A Mach-Zehnder interferometer (Fig.~\ref{fig:interferometer}) integrated into the optical path is used to characterize the mechanical properties of the nanobeams.

\section{Sample fabrication}
We start with a 4-inch, \SI{525}{\micro\meter}-thick, high resistivity ($\geq 10^4 ~\Omega \cdot \mathrm{cm}$) $\mathrm{\langle 100 \rangle}$ silicon wafer obtained from Silicon Valley Microelectronics (SVMI). These wafers have \SI{150}{\milli\meter} of high stress ($\sim 1$ GPa) silicon nitride deposited on one side using LPCVD. The backside is also scored $\sim \SI{200}{\nano\meter}$ deep according to a \SI{1}{\centi\meter} square grid.

First, we fabricate the coplanar waveguide (CPW) using optical lithography, electron-beam evaporation, and a liftoff process. The CPW consists of \SI{200}{\nano\meter} of gold evaporated on a \SI{7}{\nano\meter} sticking layer of titanium. Next, the mechanical resonators are defined with electron-beam lithography and the corresponding areas are etched with reactive ion etching, resulting in 
nanobeams that are \SI{145}{\micro\meter} long and \SI{1}{\micro\meter} wide.  Then the resonators are released using either a wet KOH etch or a dry $\mathrm{XeF_2}$ isotropic etch. Finally, the sample chip is wire-bonded to a printed circuit board (PCB) for microwave delivery.

\begin{figure}
\centering
\includegraphics[width=.6\linewidth]{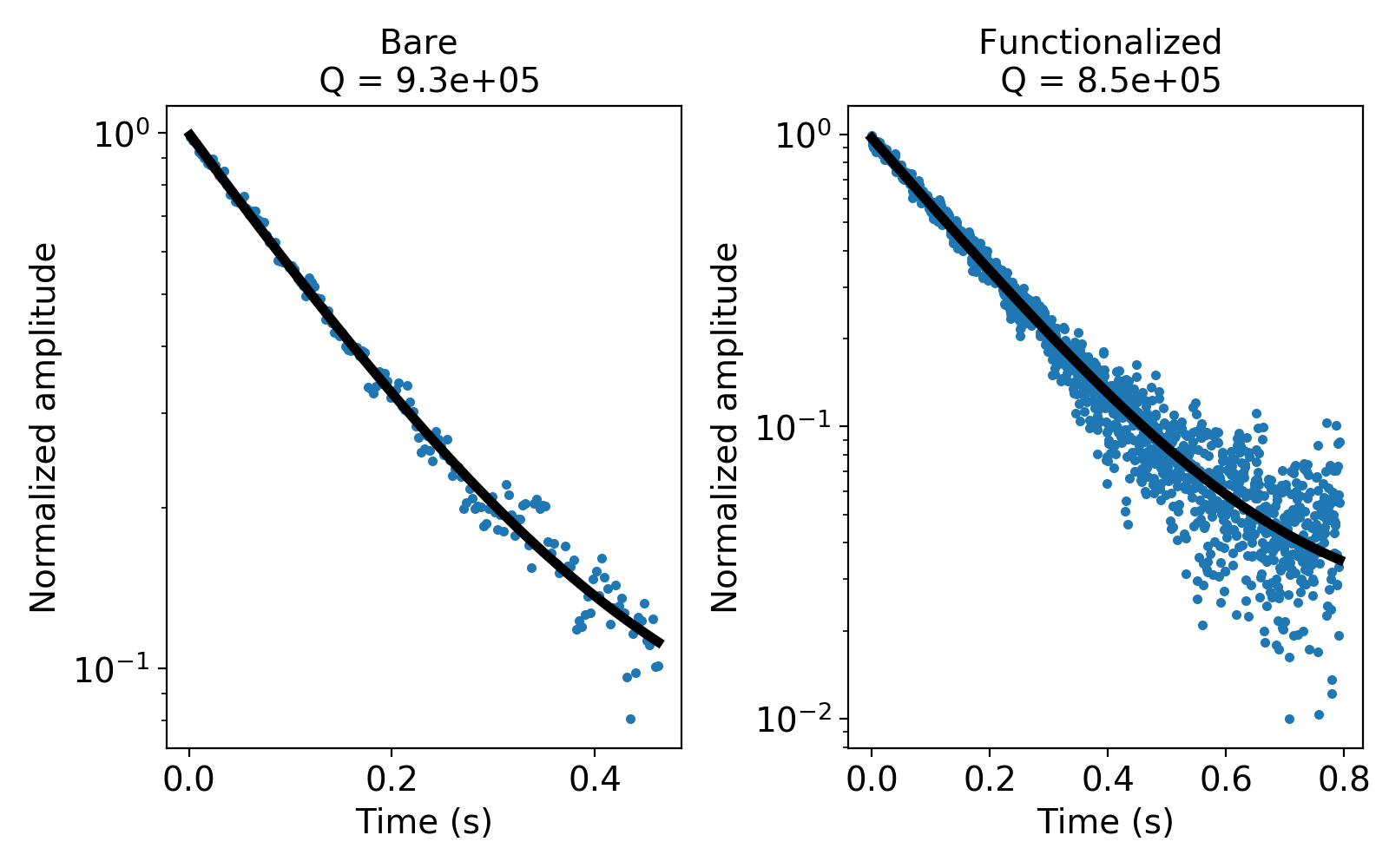}
\caption{Ringdown measurements of a bare resonator (left) and a functionalized resonator (right, same as Fig.~4(b) of the main text), taken at cryostat base temperature of \SI{20}{\kelvin} and \SI{10}{\kelvin} respectively.}
\label{fig:functionalized_Q}
\end{figure}

We magnetically functionalize the nanobeam using a home-built micromanipulator setup. The sample is placed on a 3-axis translation stage, while a tungsten tip (75960-02, Electron Microscopy Sciences) is mounted on a nearby micromanipulator. This allows us to move both the sample and tip independently under a microscope objective (100X Mitutoyo Plan Apo SL). We first deposit a small volume of UV epoxy (Loctite 349) on the center of the beam using the tungsten tip. Then we use another tungsten tip to pick up a NdFeB micromagnet, put it on the glue, and cure the epoxy with UV light. Since we have observed that the glue remains tacky even after prolonged UV exposure, we put the sample on a hot plate at $\SI{80}{\celsius}$ for $8$ hours to eliminate any ``tackiness''. A comparison of the quality factors of a functionalized and bare resonator is shown in Fig.~\ref{fig:functionalized_Q}.

Details on fabrication of the diamond nanopillars can be found in references \cite{angle_etch,zhou_scanning_2017}.

\section{Magnetic field imaging}
The scan in Fig.~2(a) in the main text was done in air at room temperature without an external field. At each position in the scan, we perform an electron spin resonance (ESR) measurement by sweeping a microwave frequency and observing changes in the NV center photoluminescence. Since we are not driving the mechanical resonator and the measurement time is much longer than the mechanical period, we can neglect the mechanics. The spin Hamiltonian is given by:

\begin{equation}
\mathcal{H}/\hbar = D S_z^2 + \gamma_e (S_z \cdot B_z + S_x \cdot B_x)
\end{equation}
where $D = \SI{2.8707}{\giga\hertz}$ is the zero-field splitting and $\gamma_e = \SI{2.8}{\mega\hertz/\gauss}$ is the NV gyromagnetic ratio. By diagonalizing $\mathcal{H}$, we can extract the magnetic field component $B_z$ from the micromagnet that is parallel to the NV center quantization axis.

Two pixels are missing because the ESR frequencies exceeded the frequency sweep range. We do not consider the missing pixels in the fit to the dipole model. For display in the figure, we fill in these pixels by interpolating from nearby pixels.

\section{Analytical form of contrast in the presence of mechanical motion}

For our analysis we consider only the $\ket{S_z = +1}$ and $\ket{S_z = 0}$ eigenstates, and model our spin-mechanical system using the following Hamiltonian:

\begin{equation}\label{Ham}
\mathcal{H}/\hbar = \frac{\omega_s}{2} \sigma_z + \omega_r a^\dagger a + \lambda 
\sigma_z (a + a^\dagger)
\end{equation}
where $a$ is the annihilation operator for the mechanical mode, $S_z = \frac{1}{2} \sigma_z$ is the spin-$\frac{1}{2}$ operator, and $\lambda = 2\pi \gamma_e z_p G$ is the spin-mechanical coupling strength for gradient $G$ and zero-point motion $z_p$. For our system, we find typical values of $z_p \sim \SI{10}{\femto\meter}$.

Considering a Hahn echo pulse sequence, we calculate the signal in the limit of no intrinsic spin decoherence. Since $\lambda$ is small, we use a semi-classical approximation, where the resonator quadrature $x = z_p (a + a^\dagger)$ is independent of the spin state.

After initializing the spin into $\ket{0}$, we put it into a coherent superposition with a $\pi/2$ pulse. We then let it evolve over time $\tau$, resulting in the following spin state, up to a global phase:

\begin{equation}
    \ket{\psi} = \frac{1}{\sqrt{2}}\left( \ket{0} + e^{-i\int_0^\tau \frac{\lambda}{z_p} x(t) dt}\ket{1} \right)
\end{equation}

After the $\pi$ pulse and evolution for another time $\tau$, the spin state becomes
\begin{equation}
\ket{\psi} = \frac{1}{\sqrt{2}}\left(\ket{0} + e^{-i \phi (\tau)}\ket{1} \right)
\end{equation}
where $ \phi(\tau) \equiv \int_\tau^{2\tau} \frac{\lambda}{z_p} x(t) dt - \int_0^\tau \frac{\lambda}{z_p} x(t) dt$ is the phase difference accumulated due to the mechanical motion. After the last $\pi/2$ pulse to convert the coherence into population, the measured contrast is $\alpha \cos{(\phi(\tau))}$, where $\alpha$ is a constant determined by the spin-dependent optical initialization and readout, as well as the background fluorescence. We assume that the quality factor is high, $Q \gg 1$, such that the signal is highly correlated during the pulse sequence, and can be given by a specific point in phase space with amplitude $x_0$ and phase $\phi_0$, such that $x(t) = x_0 \cos{(\omega_r t + \phi_0)}$. For a given run of the pulse sequence with phase $\phi_0$ and amplitude $x_0$ the accumulated phase on the spin is: 
\begin{equation}
    \phi(\tau) = \frac{\lambda}{z_p \omega_r} x_0 \left(\sin{(2\omega_r \tau + \phi_0)} - 2\sin{(\omega_r \tau + \phi_0)} + \sin{(\phi_0)} \right)
\end{equation}

The resonator frequency drifts over the duration of the Hahn echo measurement, so we apply a wide-band drive to ensure that the resonator is always driven on resonance. Because the wide-band drive effectively raises the temperature of the mechanical mode and the measurement is longer than the dissipation time,  we incoherently average over the statistical distribution for $x(t)$. From the statistics of thermal states, we assume that $\phi_0$ is uniformly distributed, and that the energy $E$ of the resonator follows a Boltzmann distribution:
\begin{equation}
p(E) dE = \frac{1}{kT} e^{\frac{E}{kT}} dE
\end{equation}
where $E = \frac{1}{2} m_{\mathrm{eff}} \omega_r^2 x_0^2$ and $\frac{1}{2} k_B T = \frac{1}{2} m_{\mathrm{eff}} \omega_r^2 \Delta_x^2$ according to the equipartition theorem; $m_{\mathrm{eff}}$ is the effective mass of the resonator, $k_B$ is the Botlzmann constant, $T$ is the effective temperature of the mechanical mode, and $\Delta_x^2$ is the variance of the mechanical motion.

%the square of the amplitude follows a Boltzmann distribution with mean $2\Delta_x^2$ (multiplied by two for each of the quadratures). %square of the amplitude $\mathcal{E}$ follows the Boltzmann exponential distribution, with mean given by $m \omega_r^2 \Delta_x^2$. 

Averaging over the phase $\phi_0$ first, we find the equivalent signal under a coherent drive with amplitude $x_0$:
\begin{equation}
    \alpha \langle \cos{(\phi(\tau))} \rangle_{\phi_0} = \alpha \mathcal{J}_0 \left(\frac{2\lambda x_0}{z_p\omega_r} \left(\cos{(\omega_r t)}-1\right)\right),
\end{equation}
where $\mathcal{J}_0$ is the zeroth order Bessel function of the first kind. Note that this would be the form of the signal under a coherent state assumption. Averaging over the amplitudes:
\begin{equation}
      \alpha \langle \cos{(\phi(\tau))} \rangle_{A, \phi_0} =\alpha \int_0^\infty \frac{1}{\Delta_x^2} e^{-x_0^2/2\Delta_x^2} \langle \cos{(\phi(\tau))} \rangle_{\phi_0} x_0 \text{ d}x_0 = \alpha e^{-q(\tau)}
\end{equation}
where $q(\tau) = 8 \Delta_x^2 \lambda^2 \sin^4{(\omega_r \tau/2)}/(\omega_r^2 z_p^2)$ as reported in the main text \cite{bennett_measuring_2012}. To account for spin decoherence, we include an additional phenomenological term $e^{\chi(\tau)}$ and factor it out with a baseline Hahn echo measurement in the absence of mechanical driving.

%As a consistency check, integrating over Gaussian distributions of the quadratures $x_0$ and $p_0 = m\omega_r \dot{x}_0$ yields the same result.

%\subsection{Consistency check 1}
%We can arrive at the same result if instead we model the position of the oscillator as $x(t) = x_0 \cos{(\omega_r t)} + \frac{p_0}{\omega m} \sin{(\omega_r t)}$ and take Gaussian averages over $x_0$ and $p_0$. YOU GET THE SAME RESULT IN THE END :)

%\subsection{Consistency check 2} COMPARING TO BENNETT ET AL: THEY INCLUDE THE EFFECT OF FINITE Q SINCE THEY DO LONGER DYNAMICAL DECOUPLING AND HAVE WORSE Q. THIS RESULTS IN NUMERICS. IN A LIMIT THEY GET AN EXPONENTIAL DECAY FORM LIKE WE DO, AND THEIR CURVES LOOK QUALITATIVELY THE SAME AS US.

\section{Preservation of coherent information while moving inside a field gradient}

\subsection{Experimental setup}
The measurements in Fig.~3 in the main text were performed in air at room temperature. An external field of about \SI{180}{\gauss} from a bar magnet is aligned within $1^{\circ}$ to the quantization axis of the NV center. To drive radiofrequency (RF) pulses on nuclear spins, we use an external coil consisting of magnet wire wound in two layers with $7$ turns each.

The sample is placed on a 3-axis Attocube stack, which in turn is placed on top of a single-axis flexure stage (Thorlabs NFL5DP20S). To move the sample containing the mechanical resonators horizontally relative to the nanopillar during the movement sequence, we drive the flexure stage with a signal from an arbitrary function generator (Tektronix AFG3021) amplified by a piezo driver (Thorlabs BPC301). The signal is chosen to be a simple sinusoid to avoid exciting higher harmonics of the setup (e.g. resonances of the Attocube stack).

\subsection{Nuclear and electronic spin control}
The RF signals are synthesized with an arbitrary function generator (Tektronix AFG3022) and amplified (Mini Circuits LZY-22+). With the RF coil placed between the nanopillar and long working distance objective (LU Plan 100X/0.90), we obtain minimum $\pi$-pulse durations of $\sim\SI{20}{\micro \second}$.

\begin{figure}
\centering
\includegraphics[width=.65\linewidth]{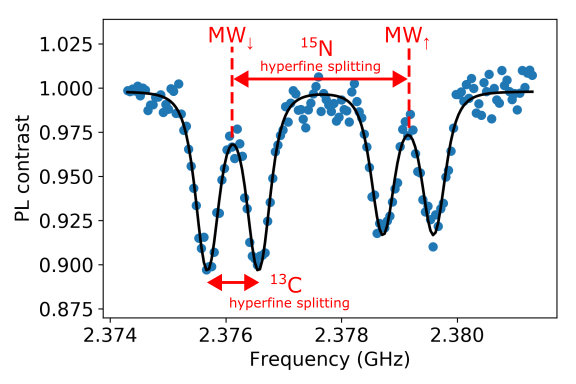}
\caption{Pulsed ESR measurement taken with the NV center being near a stationary micromagnet and an external field. We see four dips as a result of hyperfine interactions with both the $\fifteenN$ and $\thirteenC$ nuclear spins. We choose frequencies $\mathrm{MW}_\downarrow$ and $\mathrm{MW}_\uparrow$ such that they are equally detuned from the two nearby hyperfine transitions attributed to a nearby $\thirteenC$ nuclear spin.}
\label{fig:pulsed_esr_stationary}
\end{figure}

The same on-chip coplanar waveguide (CPW) in Fig.~1 of the main text is used to drive microwave (MW) transitions of the electronic spin. We generate two MW signals at the fixed frequencies $\mathrm{MW}_\uparrow$ and $\mathrm{MW}_{\downarrow}$, which correspond to the upper and lower $\fifteenN$ hyperfine transitions (Fig.~\ref{fig:pulsed_esr_stationary}). The two MW signals are combined using a power combiner, gated with a switch, and then amplified (Mini Circuits ZHL 16W 43+).

During the movement sequence, the micromagnet begins near the NV center, moves away, and finally returns to its original position. The pulse sequences for the coherence measurements are synchronized with the movement sequence. While the hyperfine frequencies shift significantly over the entire movement sequence, we expect from the sinusoidal driving signal that the shift will be insignificant during the beginning and end, which are the only times where $\CnNOTe$ pulses are applied. We estimate from our field profile measurement in Fig.~3(a) that during the time between the two entangling/disentangling gates, the hyperfine frequencies do not deviate by more than \SI{15}{\kilo\hertz}, corresponding to an estimated phase accumulation of less than $0.01\,\pi$ over a maximum accumulation time of $\SI{4.5}{\micro\second}$.

\subsection{Pulse sequence for ESR measurements during movement sequence}

\begin{figure}
\centering
\includegraphics[width=.5\linewidth]{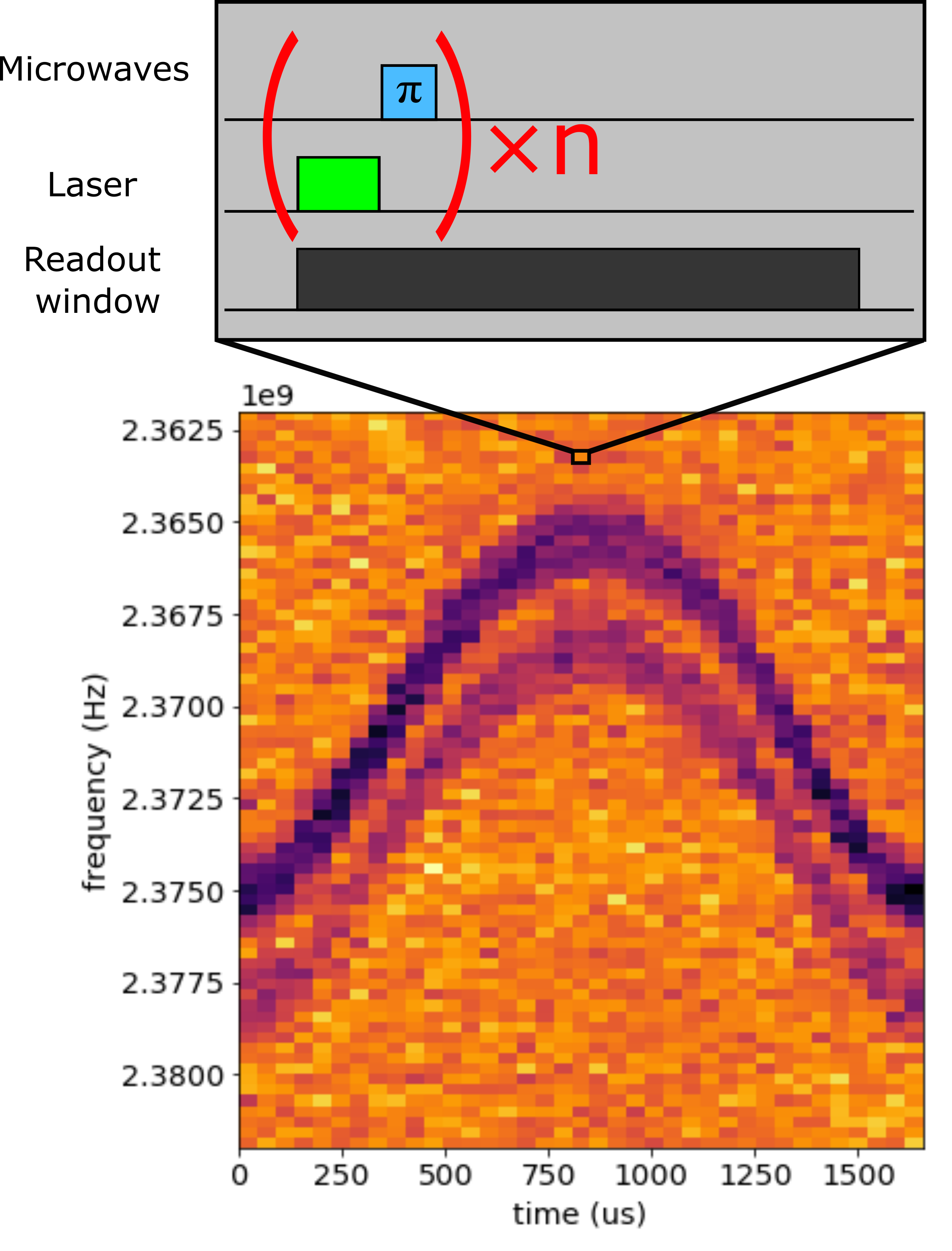}
\caption{Pulse sequence for taking ESR measurements during movement. For each microwave frequency, we divide the total movement time by a number of time segments (number of pixels along the x-axis). In each time segment we run the pulse sequence shown inside the gray box.}
\label{fig:pulsed_esr_moving}
\end{figure}

To monitor the field experienced by the NV center during the movement of the micromagnet (Fig.~3(a) in main text), we take pulsed ESR measurements at different times and MW frequencies during the movement sequence (Fig.~\ref{fig:pulsed_esr_moving}). By initializing the NV center electronic spin and then applying a $\pi$ pulse, we expect to see a drop in the NV photoluminescence if the MW frequency coincides with the ESR frequency. To obtain a better signal-to-noise ratio, we repeat for $n = 11$ times the initialization and $\pi$-pulse pair, for each time segment ($\sim \SI{46}{\micro\second}$) of the entire movement duration of \SI{1.7}{\milli\second}.

Due to the long read times of our data acquisition card, we use a single long readout window for each time segment, which results in faster data acquisition at the expense of lower photoluminescence contrast, as the readout window now includes times where the NV center is being reinitialized.

\subsection{Pulse sequence for coherence measurement}
\begin{figure}
\centering
\includegraphics[width=.9\linewidth]{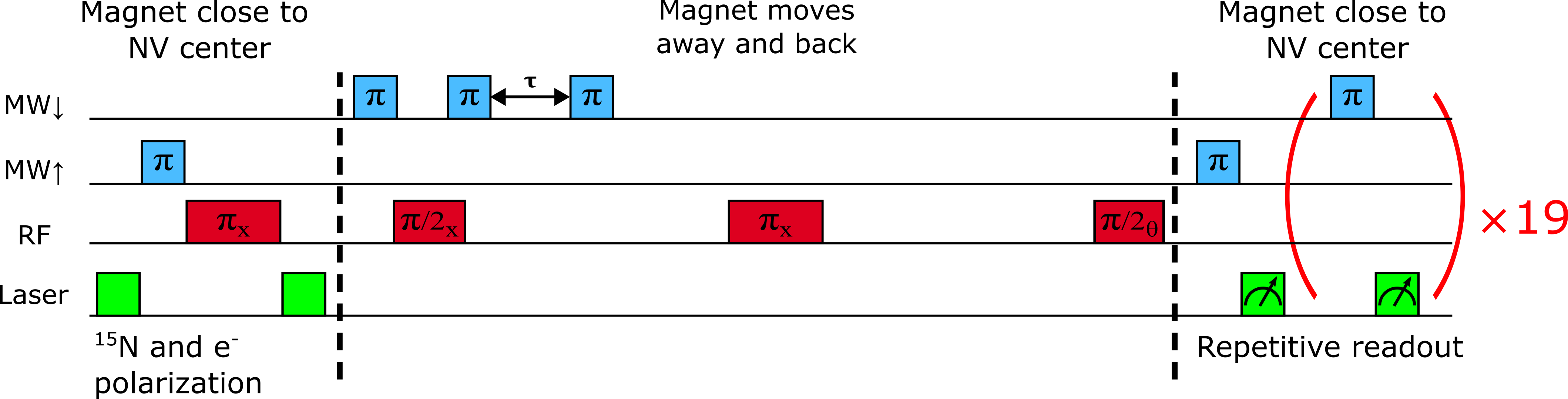}
\caption{Full pulse sequence for the measurements in Fig.~3(c) in the main text, with the part between the dashed lines corresponding to the pulse sequence in Fig.~3(b).}
\label{fig:full_pulse_sequence}
\end{figure}

\begin{figure}
\centering
\includegraphics[width=0.6\linewidth]{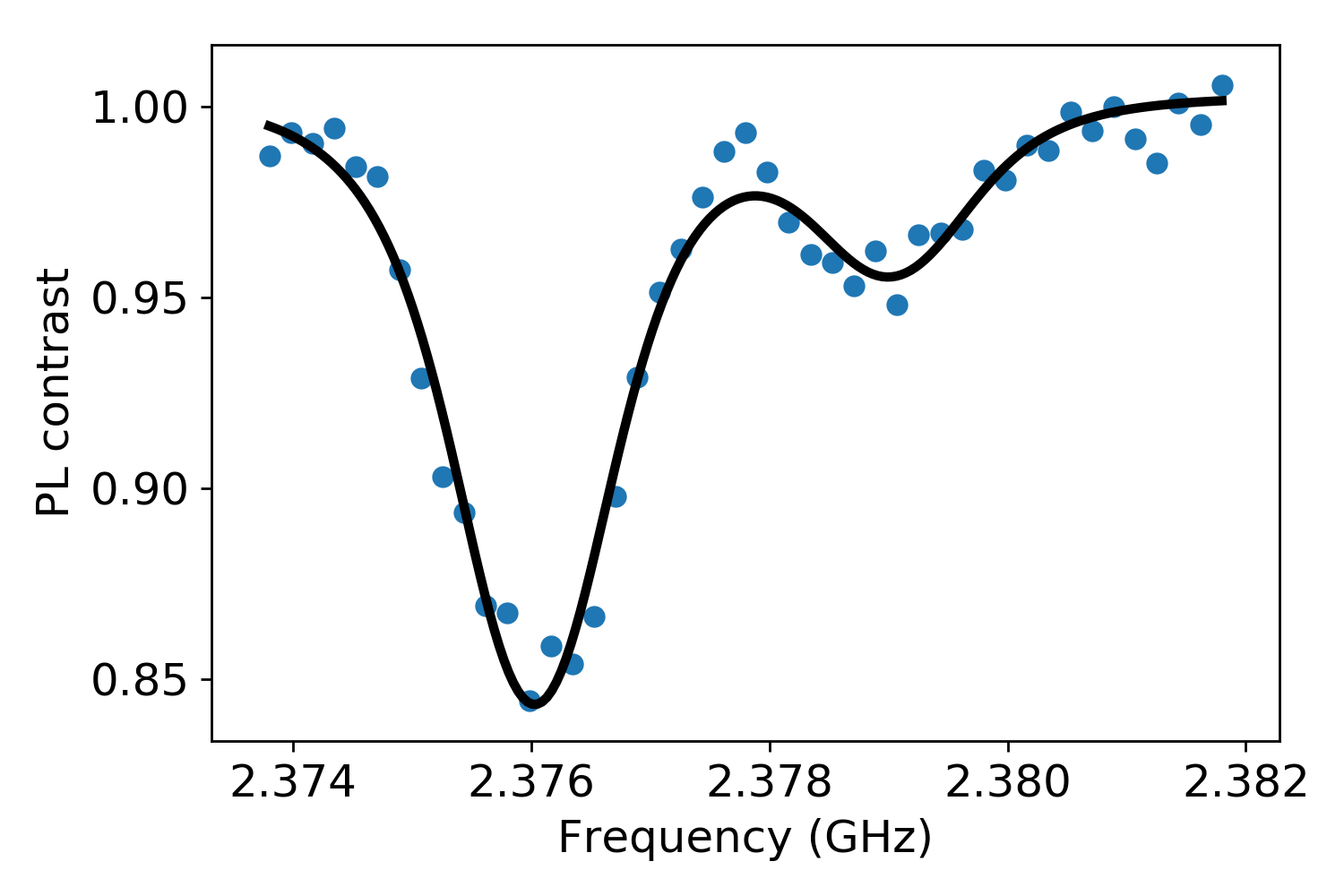}
\caption{Pulsed ESR taken after nuclear spin polarization. }
\label{fig:nuclear_polarization}
\end{figure}

The detailed version of the pulse sequence in Fig.~3(b) in the main text is shown in Fig.~\ref{fig:full_pulse_sequence}. At the beginning of the pulse sequence, we polarize the $\fifteenN$ nuclear spin by first polarizing the electron spin and then transferring the polarization to the nuclear spin via a SWAP operation \cite{jiang2009repetitive}. A pulsed ESR measurement after $\fifteenN$ polarization is shown in Fig.~\ref{fig:nuclear_polarization}, where the polarization is $\sim\SI{78}{\percent}$.

To normalize the measurement in Fig.~3(c) in the main text, we use the same pulse sequence, but without the two entangling and disentangling $\CnNOTe$ gates. Setting the rotation axis angle of the final nuclear $\pi/2$ pulse to be $\theta = 0$ and $\pi$, we obtain the upper and lower bounds for the fluorescence levels, which we use to normalize the measurements where we vary $\theta$.

Despite the small gyromagnetic ratio of the the $\fifteenN$ nuclear spin, the field change during the movement sequence is large enough to cause significant phase accumulation ($>2 \pi$) on the nuclear spin. To cancel out this phase, we apply a $\pi$-pulse on the $\fifteenN$ nuclear spin. Since the field changes symmetrically during the movement sequence, the optimal time for this $\pi$-pulse is around the middle of the movement sequence, when the micromagnet is the furthest away from the NV center.

One can calculate the optimal $T_\pi$ by measuring the ESR frequencies of the electronic spin during the movement sequence (Fig.~3(a) in the main text). As the micromagnet moves away from the diamond nanopillar, the ESR frequencies become positively detuned relative to their initial values due to the Zeeman interaction; as the micromagnet moves back, the detunings of the ESR frequencies decrease back to $0$. The nuclear magnetic resonance (NMR) frequency of the $\fifteenN$ nuclear spin also experiences the same detuning as the electron spin, scaled by the ratio of their respective gyromagnetic ratios $\gamma_{\mathrm{n}}/\gamma_e$. We then solve for $T_\pi$ such that the total phase accumulated $\phi_{\mathrm{n}}$ is $0$:

\begin{equation}
\phi_{\mathrm{n}} = \int^{T_\pi}_0 \gamma_{\mathrm{n}}\delta_{\mathrm{n}}(t) dt - \gamma_{\mathrm{n}} \int^{T_{\mathrm{move}}}_{T_\pi} \gamma_{\mathrm{n}} \delta_{\mathrm{n}}(t) dt = 0
\label{eqn:phase_accum}
\end{equation}
where $T_{\mathrm{move}} = \SI{1.7}{\milli\second}$ is the duration of the entire movement sequence; $\delta_n(t)$ is the time-dependent detuning of the $\fifteenN$ NMR frequency from its frequency at the beginning of the movement sequence.

\begin{figure}
\centering
\includegraphics[width=.65\linewidth]{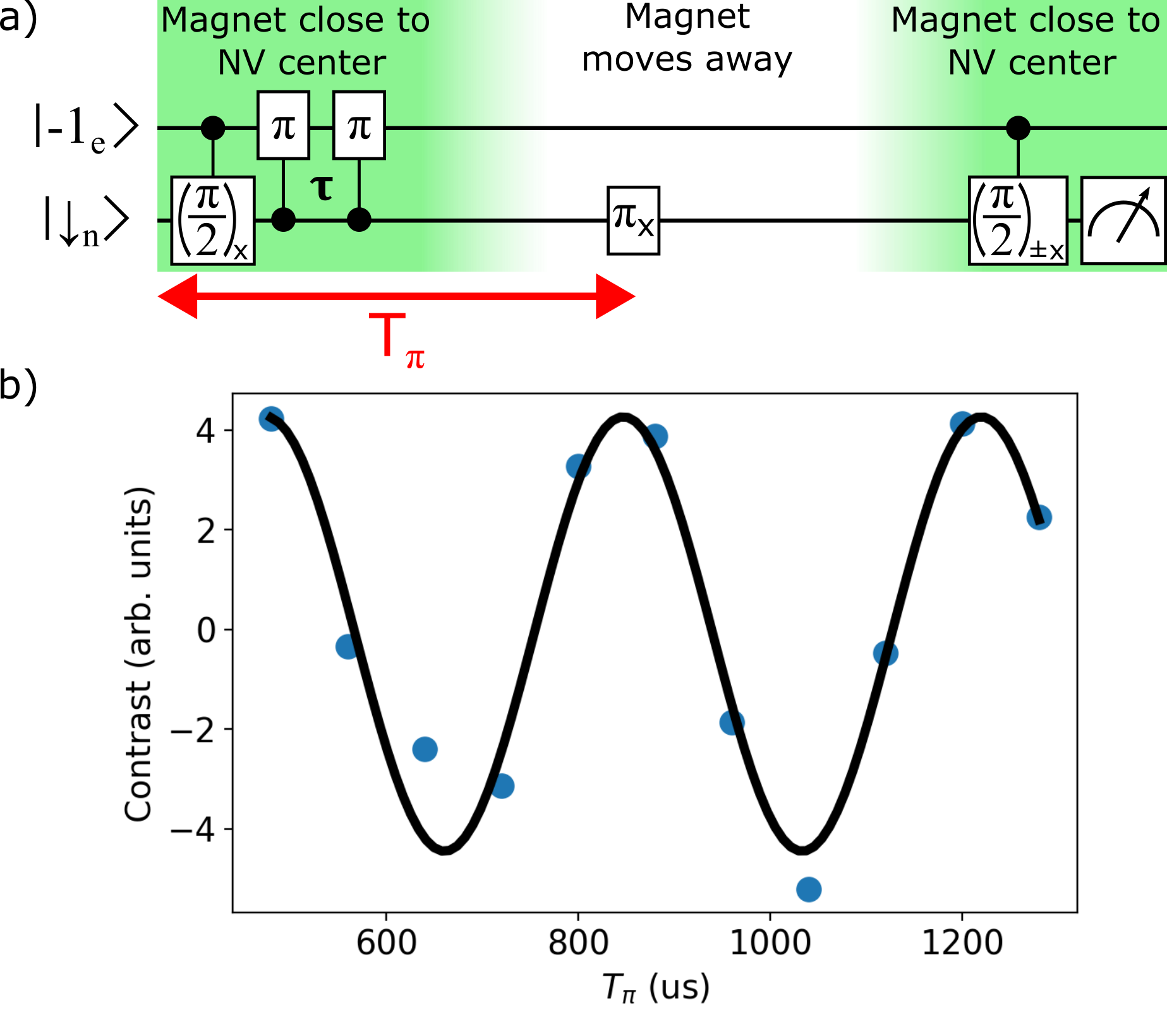}
\caption{(a) Pulse sequence for determining the optimal time of the $\fifteenN$ nuclear spin echo $\pi$-time. (b) We observe fringes in the readout contrast as we sweep $T_\pi$, where the maxima correspond to times when the phase accumulation is cancelled out to $0$ by the nuclear $\pi$-pulse. The contrast is defined as the difference between the readout fluorescence when the final $\pi/2$-pulse rotation axis is set to $+x$ and $-x$.}
\label{fig:vary_pi_time}
\end{figure}

To experimentally determine $T_\pi$, we use the pulse sequence in Fig.~\ref{fig:vary_pi_time}, where we vary the $\pi$-pulse time $T_\pi$ and observe the resulting change in readout photoluminescence. As expected, as we vary $T_\pi$, the total phase accumulation also changes, resulting in the fringes in Fig.~\ref{fig:vary_pi_time}. The maxima of the fringe correspond to the times when the total phase accumulation is $0$. The maximum at $\sim \SI{853}{\micro\second}$ is roughly half the time of the movement sequence and agrees with the calculated result using Equation~\ref{eqn:phase_accum}.

\subsection{Phase information $\phi(\tau)$}
\begin{figure}
\centering
\includegraphics[width=.65\linewidth]{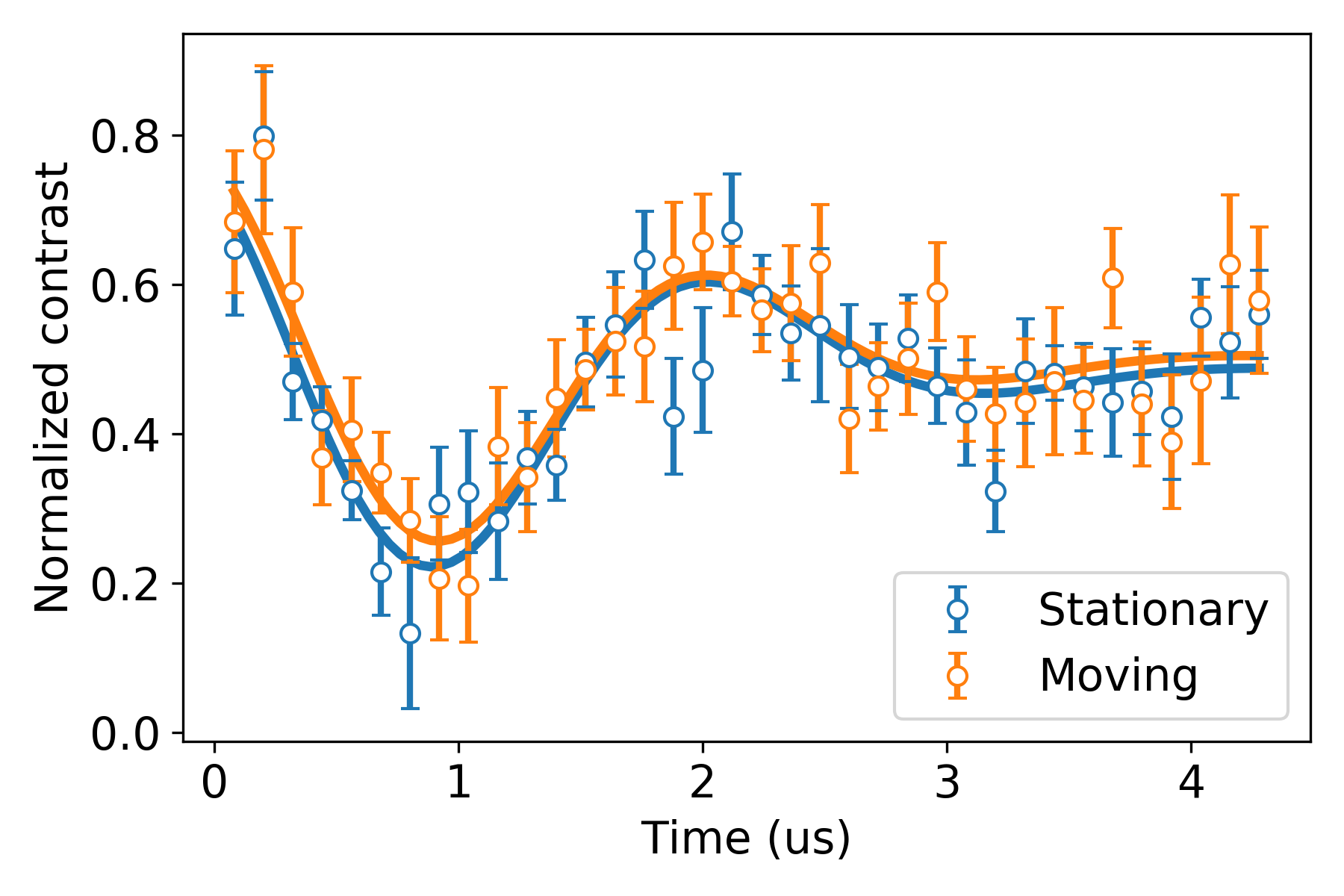}
\caption{Phase information stored inside the nuclear spin. The fringe contrast is not degraded by movement inside the magnetic field gradient.}
\label{fig:phase_vary_tau}
\end{figure}

In fig.~3(c) of the main text, we quantify the preservation of spin coherence by fixing $\tau = \SI{900}{\nano\second}$ and varying the rotation axis angle of the final $\pi/2$ pulse. We also show in Fig.~\ref{fig:phase_vary_tau} a measurement where we keep the rotation axis angle of the final $\pi/2$ pulse same as all the other nuclear pulses, and vary $\tau$ instead. As expected, we observe Ramsey fringes due to the $\CnNOTe$ gate having an equal detuning from the two $\thirteenC$ transitions. 

\section{NV center electronic and nuclear spin coherence times}

\begin{figure}
\centering
\includegraphics[width=1.0\linewidth]{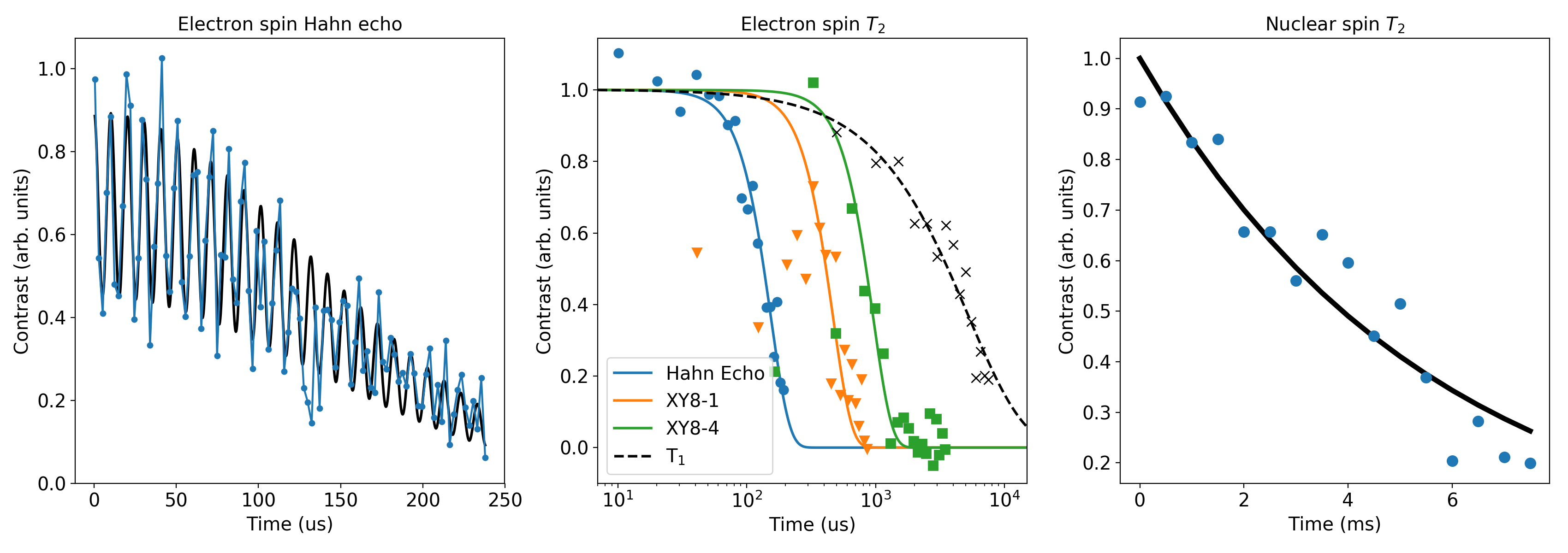}
\caption{Left: coherence of a single NV in the nanopillar, measured using a Hahn echo sequence, showing collapse and revivals of the 13C bath due to an external bias magnetic field. Middle: coherence times measured using different dynamic decoupling sequences. The electron spin $T_1$ is also plotted (black) for comparison. Right: coherence of the nuclear spin, measured using a Hahn echo sequence.}
\label{fig:t1_t2}
\end{figure}

In Fig.~\ref{fig:t1_t2}, we show measurements of the electron spin coherence time ($T_{2,e}$), lifetime ($T_{1,e}$), and the $\fifteenN$ nuclear spin coherence time ($T_{2,n}$). These measurements are performed in air at room temperature in the presence of an external bias magnetic field of $\sim \SI{180}{\gauss}$. With an XY8-4 sequence, $T_{2,e}$ exceeds \SI{0.95(4)}{\milli\second} (middle). We expect that NV centers implanted with more energy than the 6 keV used in this diamond nanopillar would have improved coherence times. Furthermore, we measure the nuclear spin coherence time $T_{2,n} \sim \SI{5}{\milli\second}$ (right), similar to $T_{1,e}$, suggesting that it is limited by the electron spin lifetime.

\section{Comparisons of Cooperativity for Previous Works}
In Table~\ref{table:cooperativities} we compare the spin-mechanical cooperativity of our current system with those of previous works involving NV centers. We note that some of the previous works do not explicitly report parameters necessary to calculate the cooperativity; we give optimistic estimates (highlighted) where appropriate.

\begin{table}[]
\centering
\begin{tabular}{lcccccc}
\hline
Work &
  \begin{tabular}[c]{@{}c@{}}$\omega/(2\pi)$ \\ (Hz)\end{tabular} &
  \begin{tabular}[c]{@{}c@{}}$g/(2\pi)$\\ (Hz)\end{tabular} &
  \begin{tabular}[c]{@{}c@{}}T\\ (K)\end{tabular} &
  \begin{tabular}[c]{@{}c@{}}$n \kappa/(2\pi)$\\ (Hz)\end{tabular} &
  \begin{tabular}[c]{@{}c@{}}$T_2$\\ (s)\end{tabular} &
  $C$ \\ \hline
This work        & $\num{1.4e6}$ & $\num{7.7}$    & 20  & $\num{5e5}$  & $\num{8.8e-4}$ & $\num{1.0e-7}$  \\
Gieseler \textit{et al.} \cite{gieseler_single-spin_2020} & $\num{1.4e2}$ & $\num{4.8e-2}$ & \colorbox{yellow}{4}   & $\num{8e4}$  & $\colorbox{yellow}{\num{1.0e-2}}$ & $\num{2.8e-10}$ \\
Arcizet \textit{et al.} \cite{arcizet_single_2011} & $\num{1.0e6}$ & $\num{1.4e2}$  & 300 & $\num{4e10}$ & $\num{4.0e-4}$ & $\num{1.8e-10}$ \\
Oeckinghaus \textit{et al.} \cite{wrachtrup_cantilever_2020} & $\num{8.6e5}$ & $\num{7.7}$    & 5   & $\num{2e5}$  & $\num{1.2e-6}$ & $\num{2.8e-10}$ \\ \hline
\end{tabular}
\caption{Comparison of spin-mechanical cooperativities with previous works. Optimistic estimates (highlighted) are given for unreported values. We also note the large coupling strength ($> \SI{700}{\hertz}$ reported in a related work where a cantilever is coupled to unpaired electron spins ($\sim \SI{700}{\hertz})$ \cite{rugar_single_2004}; however, the spin coherence time and quality factor were not reported.}
\label{table:cooperativities}
\end{table}

\bibliography{SI_bib}